\documentclass[aps,prb,superscriptaddress,floatfix,showpacs,twocolumn]{revtex4}
\usepackage{amsfonts,amsmath,amssymb,bm}
\usepackage{graphicx,graphics,float}
\usepackage{color}
\usepackage[colorlinks,citecolor=blue,linkcolor=blue]{hyperref}
\usepackage{comment}

\newcommand{\be}{\begin{equation}}
\newcommand{\ee}{\end{equation}}
\newcommand{\bea}{\begin{eqnarray}}
\newcommand{\eea}{\end{eqnarray}}

\newcommand{\vep}{\varepsilon}

\newcommand{\ome}{\omega}

\begin{document}
\title{Quantum-dot circuit-QED thermoelectric diodes and transistors}

\author{Jincheng Lu}
\address{School of Physical Science and Technology \&
Collaborative Innovation Center of Suzhou Nano Science and Technology, Soochow University, Suzhou 215006, China.}
\address{International Centre for Theoretical Sciences, Tata Institute of Fundamental Research, Bangalore 560089, India}

\author{Rongqian Wang}
\address{School of Physical Science and Technology \&
Collaborative Innovation Center of Suzhou Nano Science and Technology, Soochow University, Suzhou 215006, China.}

\author{Jie Ren}
\address{Center for Phononics and Thermal Energy Science, School of Physical Science and Engineering, Tongji University, 200092 Shanghai, China }

\author{Manas Kulkarni }\email{manas.kulkarni@icts.res.in}
\address{International Centre for Theoretical Sciences, Tata Institute of Fundamental Research, Bangalore 560089, India}

\author{Jian-Hua Jiang}\email{jianhuajiang@suda.edu.cn, joejhjiang@hotmail.com}
\address{School of Physical Science and Technology \&
Collaborative Innovation Center of Suzhou Nano Science and Technology, Soochow University, Suzhou 215006, China.}

\date{\today}
\begin{abstract}
Recent breakthroughs in quantum-dot circuit-quantum-electrodynamics (circuit-QED) systems are
important both from a fundamental perspective and from the point of view of quantum photonic devices.
However, understanding the applications of such setups as potential thermoelectric diodes and transistors has
been missing. In this paper, via the Keldysh nonequilibrium Green's function approach, we show that cavity-coupled double quantum-dots can serve as excellent quantum thermoelectric diodes and transistors. Using an enhanced perturbation approach based on polaron-transformations, we find non-monotonic dependences of thermoelectric transport properties on the electron-photon interaction. Strong light-matter interaction leads to pronounced rectification effects for both charge and heat, as well as thermal transistor effects in the linear transport regime, which opens up a cutting-edge frontier for quantum thermoelectric devices.
\end{abstract}


\maketitle

\section{Introduction}
Recently, there has been a flurry of activities and progress in probing and controlling hybrid light-matter systems which sit at the confluence of mesoscopic physics and quantum  optics\cite{RenRMP,RevNori,nat_phys_rev,Nanotechnology,JiangCRP,THIERSCHMANN20161109,BENENTI20171}. Few examples of such hybrid light-matter systems include quantum-dot (QD) circuit-Quantum Electrodynamics (c-QED)  systems\cite{PhysRevA.90.062305,PhysRevB.90.125402,PhysRevX.6.011032,Fluorescence,PhysRevLett.107.256804,PhysRevLett.114.196802,Petta1,Mi156,PhysRevLett.113.036801}, cold atoms coupled to light\cite{PhysRevLett.111.220408}, and optomechanical devices\cite{PhysRevLett.109.253601,PhysRevLett.108.153603,PhysRevLett.110.253601,RevModPhys.86.1391,peng2014parity,chang2014parity,Nat.Phys}.
{Rich emergent quantum phenomena have been found in recent experiments} where QDs at finite voltage bias have been integrated with superconducting microwave resonators\cite{PhysRevApplied.6.054013,Liu285,PhysRevLett.107.256804,kontos15,kontosNatt}, accomplishing sufficiently strong light-matter coupling. Such QD-cQED systems offer a rich platform for studying nonequilibrium open quantum systems. Experiments are versatile, tunable (large windows of parameters) and scalable. These QD c-QED setups are important both from a fundamental perspective (investigating correlations, transport, entanglement, bosonic statistics) and from the point of view of device applications (quantum microwave amplifiers and lasers in microwave regime). From the perspective of devices, the focus and success until now has been on realizing photon emitters, microwave amplifiers and even single-atom lasers \cite{single-atom-petta}. However, there has been no work on investigating these systems as potential quantum {transistors}
and rectifiers \cite{RenRect,Jiangtransistors,Archak1,RenFCG}, which is the aim of this {work}.

Manipulation and separation of thermal and electrical currents at mesoscopic scales are of fundamental interest and have technological impact on high-performance thermoelectric  devices  \cite{Mahan7436,Kuo,JiangCRP,RenPRB,Jiang2012,Ruokola,Jiang2013,JiangNJP,marco,JiangJAP,JiangOraPRE,Bergenfeldt,JiangPRL,Jiangtransistors,S2015Heat,Thierschmann2015Thermal,BijayJiang,JiangSR,Agarwalla2015,JiangPRApplied,S2017Single,JiangNearfield,Rafael1,MyJAP,Mecklenburg,Rongqian,Tang,ManasPermanent,Rafael2}. In this paper, we investigate the inelastic thermoelectric transport assisted by microwave photons residing in the cavity, as well as elastic tunneling transport. The strong light-matter interaction provides an excellent avenue for realizing quantum thermoelectric devices. By employing the non-equilibrium Green's function approach\cite{Ora3,Wang2014,BijayManas1,BijayManas2,GF1,GF2,GF3}, we show that due to the nonlinearity induced
by the electron-photon interaction, significant charge and thermal rectification effects can be realized by properly tuning the QDs energy. We further show that these QD c-QED setups exhibit thermal transistor effects even in the linear transport regime, and thus provide a {salient} platform for unprecedented thermal control.

\begin{figure}[htb]
  \centering \includegraphics[width=4.2cm]{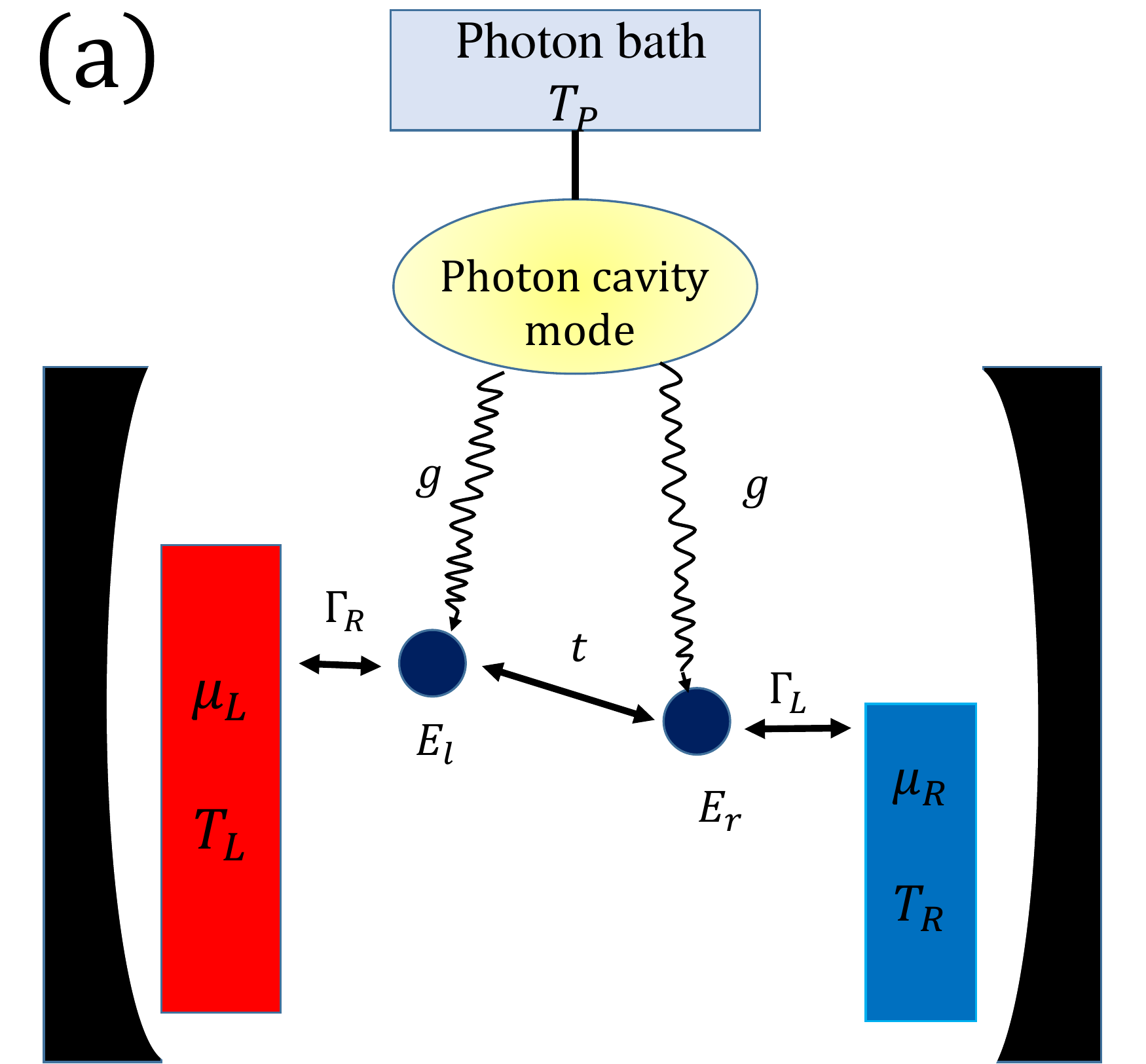}\hspace{0.2cm}\includegraphics[width=4.2cm]{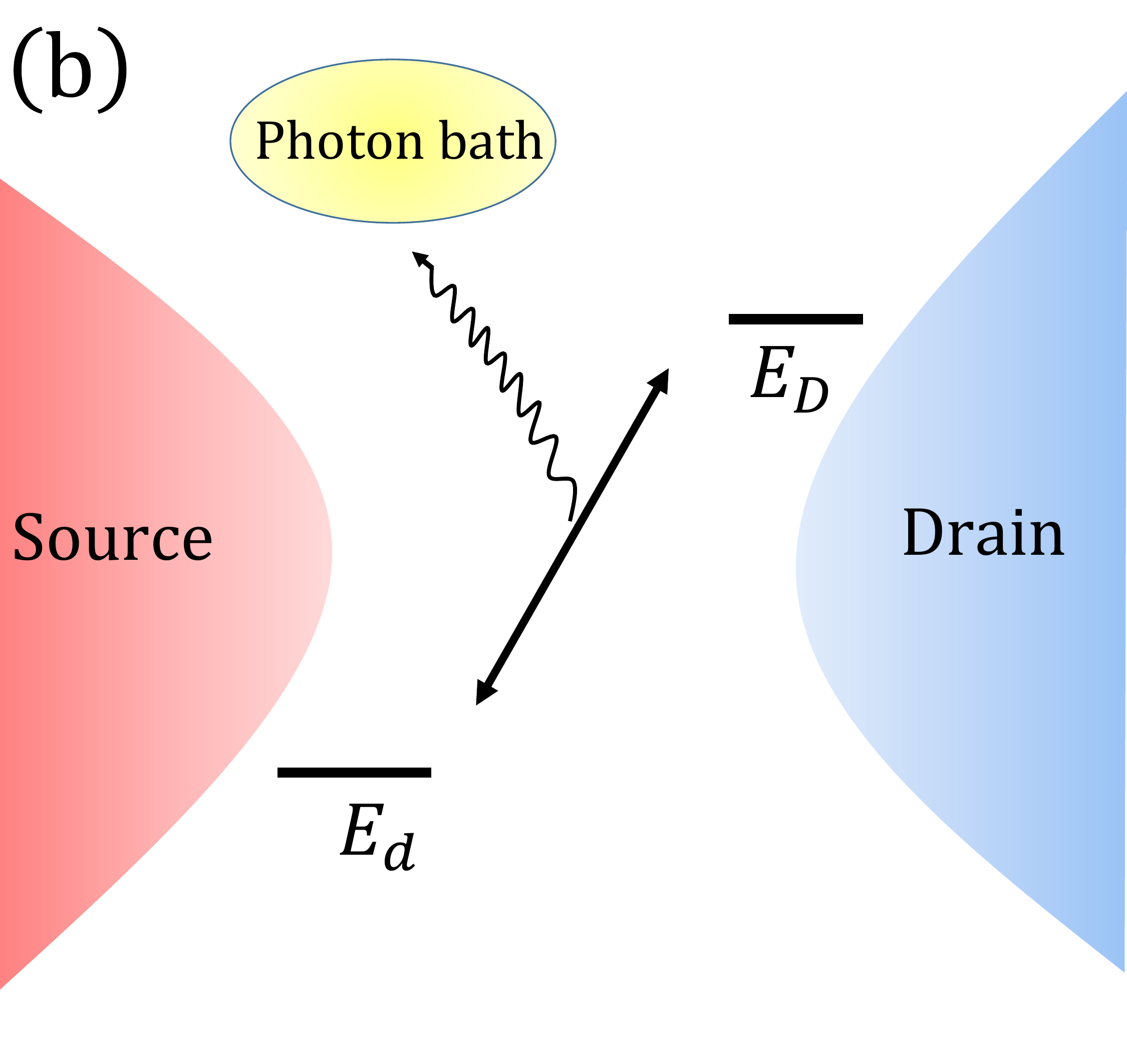}
  \centering \includegraphics[width=4.2cm]{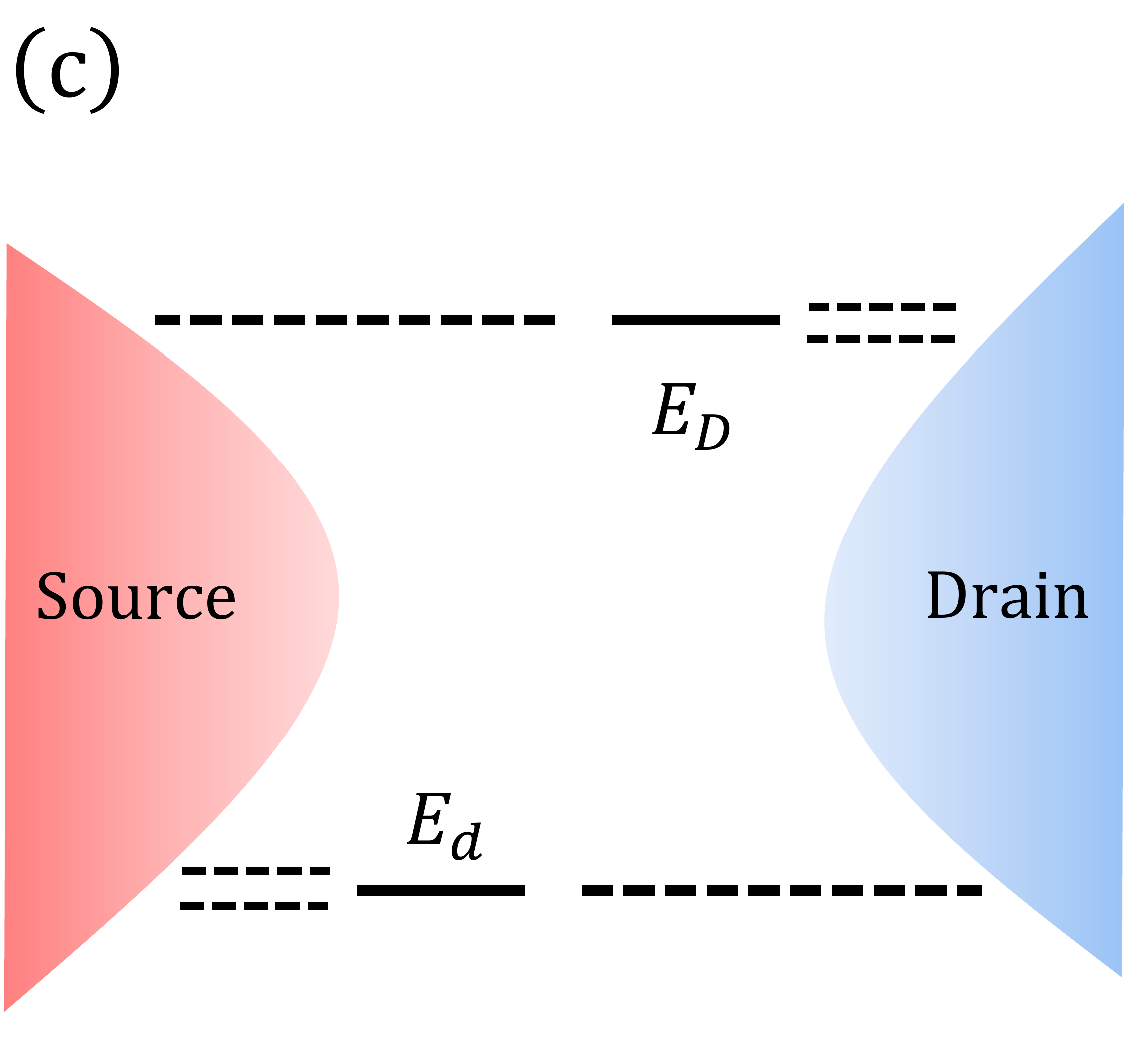}\hspace{0.2cm}\includegraphics[width=4.2cm]{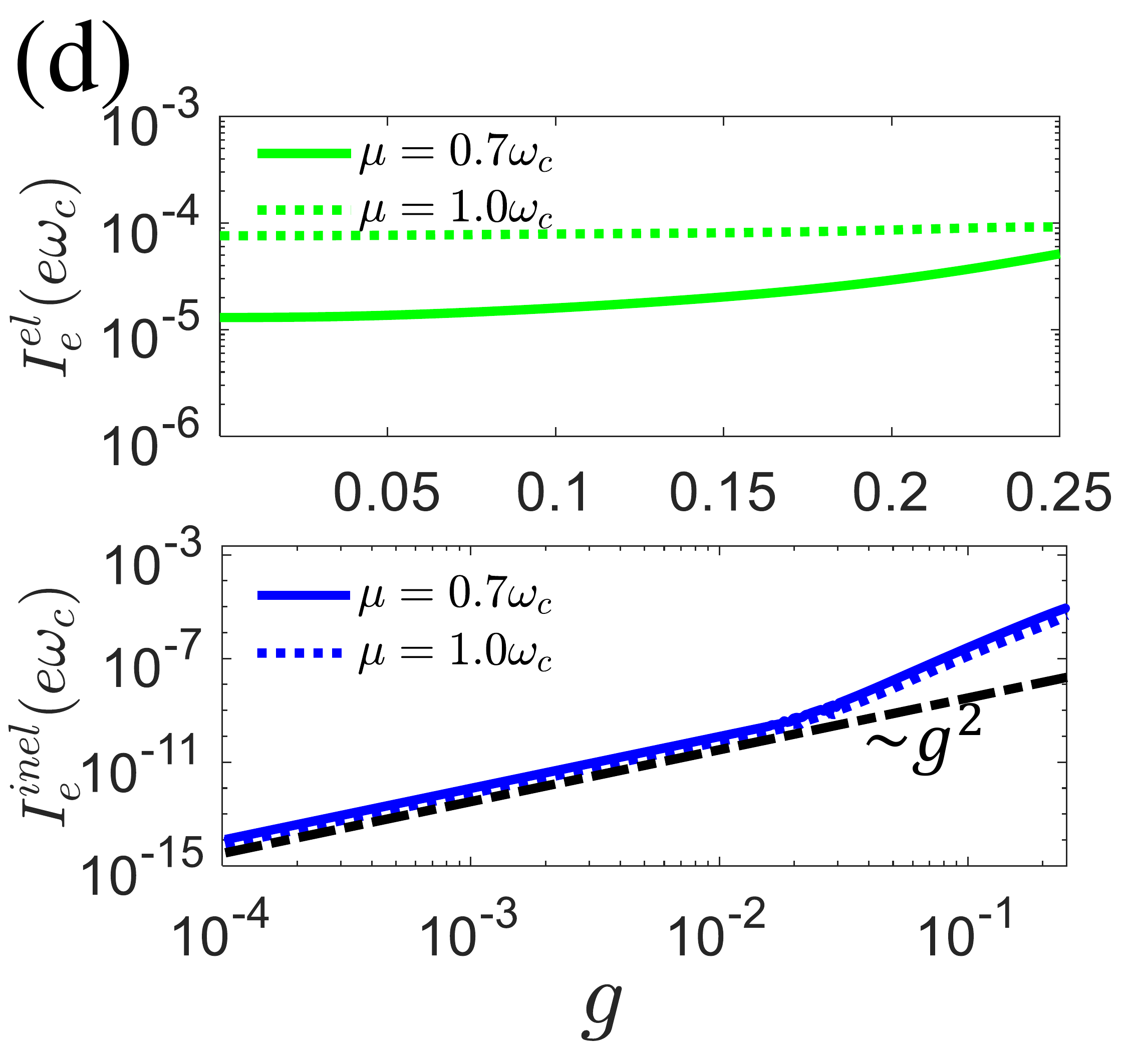}
 \caption{(Color online) (a) A schematic representation of the model. The mesoscopic system is effectively housed in the microwave cavity. Wavy lines indicate the light-matter coupling $g$. Tunneling rates between the dots and the electron leads ($\Gamma_L$, $\Gamma_R$) and in between the dots ($t$) can be tuned via gate-controlled tunnel barriers. Electrons travel from source into the first QD (with energy $E_l$) and then hop to the second QD (with a different energy $E_r$) assisted by a photon from the photonic bath. (b) Illustration of possible photon-assisted inelastic transport processes. (c) Illustration of possible elastic transport processes. Here $E_d$ and $E_D$ are the QDs energy after hybridization.  (d) {Elastic ($I_e^{el}$) and inelastic ($I_e^{inel}$) electric currents as functions of the light-matter interaction constant $g$ for different chemical potentials $\mu$}, where $E_l=0$, $E_r=\omega_c$, $\Gamma_0=0.1\omega_c$,  $k_BT_L=k_BT_R=k_BT_p=k_BT=0.1\omega_c$, $t=0.3\omega_c$, and $\Delta\mu=0.01\omega_c$.}
\label{fig1}
\end{figure}

\section{Model and thermoelectric transport in cavity-coupled double quantum-dot systems}
As schematically depicted in Fig.~\ref{fig1}, we consider double QDs (DQD) that {are} connected to two electronic reservoirs and a photonic bath. The QDs are defined with tunable electronic energy levels $E_l$ and $E_r$ by local gate-voltages. $t$ is the tunneling between the QDs. $\Gamma_L$ and $\Gamma_R$ are the hybridization energies of the dots to the source and drain electrodes (labeled by $L$ and $R$, respectively), respectively. Charge current, electronic heat current, and photonic heat current are induced by applying a voltage bias between the terminals $L$ and $R$ and temperature differences {among} the three {reservoirs}. The {entire} system is described by the Hamiltonian,
\begin{equation}
\begin{aligned}
\hat H = \hat H_{c-DQD} + \hat H_{lead} + \hat H_{dot-lead},
\end{aligned}
\end{equation}
with
\begin{equation}
\begin{aligned}
\hat H_{c-DQD} = \hat H_{DQD} + \hat H_{p} + \hat H_{e-p}.
\end{aligned}
\end{equation}
Note that in the above equation, $H_{c-DQD}$  {consists of the Hamiltonian for the double QDs ($H_{DQD}$), the Hamiltonian for the single-mode cavity photon ($H_{p}$), and the interaction between them ($H_{e-p}$),} as elucidated below,
\begin{subequations}
\begin{align}
& \hat H_{DQD} = \sum_{i=\ell,r} E_i \hat d_i^\dagger \hat d_i + ( t \hat d_\ell^\dagger \hat d_r + {\rm H.c.}) , \\
& \hat H_{p} = \ome_{c}\hat a^\dagger \hat a, \\
& \hat H_{e-p} = g\omega_c \hat (d_\ell^\dagger \hat d_\ell+d_r^\dagger \hat d_r)(\hat a + \hat a^\dagger) .
\end{align}
\end{subequations}
Here $\hat d_{l/r}^\dagger$ creates an electron in the $i$th QD with an energy $E_{l/r}$. The $l(r)$ QD is located next to and strongly coupled with the left (right) lead. The tunneling elements from the $l$ QD to the right lead and that from the $r$ QD to the left lead are assumed negligible. $\hat a^\dagger$ and $\hat a$ create and annihilate a photon with energy $\omega_c$ (we set $\hbar\equiv 1$ throughout this paper) {in the single-mode cavity, respectively.} The last term describes the light-matter interactions {characterized by the 
dimensionless parameter $g$}. The Hamiltonians
\begin{subequations}
\begin{align}
&\hspace{-0.3cm} \hat H_{lead} = \sum_{j=L, R}\sum_{k} \vep_{j,k} \hat d_{j,k}^\dagger \hat d_{j,k}, \\
&\hspace{-0.3cm} \hat H_{dot-lead} = \sum_k V_{L, k} \hat d_\ell^\dagger \hat d_{L, k} + \sum_k V_{R, k}\hat d_r^\dagger \hat d_{R, k}+{\rm H.c.}
\end{align}
\end{subequations}
describe the electronic leads and the tunneling between the QDs and the leads, respectively.

We first diagonalize the {DQD Hamiltonian $\hat H_{DQD}$}, and write it in terms of a new set of electronic operators $\hat D=\sin\theta \hat d_l + \cos\theta \hat d_r$ and $\hat d=\cos\theta \hat d_l - \sin\theta \hat d_r$, where $\theta\equiv \frac{1}{2}\arctan(\frac{2t}{\epsilon})$ and $\epsilon\equiv E_r-E_l$. The corresponding levels are $E_D=\frac{E_r+E_l}{2}+\sqrt{\frac{\epsilon^2}{4}+t^2}$ and $E_d=\frac{E_r+E_l}{2}-\sqrt{\frac{\epsilon^2}{4}+t^2}$. Using these operators, {we can write the DQD c-QED Hamiltonian as}  $\hat H_{c-DQD} =  E_D\hat D^\dagger\hat D + E_d\hat d^\dagger\hat d+\omega_c\hat b^\dagger \hat b
+ g(\hat D^\dagger\hat D+\hat d^\dagger\hat d)(\hat b^\dagger + \hat b)$.
By employing {$\Gamma_j(\omega)=2\pi\sum_k{|V_{j,k}|^2\delta(\omega-\epsilon_{j,k})}$ with $j=L,R$, the tunneling rates between the leads and the QDs} (in the local basis $d_l^\dag|0\rangle$ and $d_r^\dag|0\rangle$) become
\begin{equation}
\begin{aligned}
\hat \Gamma^L =  \left( \begin{array}{cccc} \Gamma_L & 0 \\ 0 &
    0 \end{array} \right), \  \hat \Gamma^R =  \left( \begin{array}{cccc} 0 & 0 \\ 0 &
    \Gamma_R \end{array} \right).
\end{aligned}
\end{equation}
The unitary transformation matrix between the local basis and the {new} basis is $U =  \left( \begin{array}{cccc} \sin\theta & \cos\theta \\ -\cos\theta & \sin\theta \end{array} \right)$. Hence, the tunnel coupling matrices between the QDs and the leads in the {new} basis become
\begin{equation}
\begin{aligned}
& \hat \Gamma^L_{rot} = U\hat \Gamma^L U^\dagger= \Gamma_L \left( \begin{array}{cccc} \sin^2\theta &  - \cos\theta \sin\theta \\-\ \cos\theta \sin\theta &
    \cos^2\theta  \end{array} \right), \\
& \hat \Gamma^R_{rot} = U\hat \Gamma^R U^\dagger=  \Gamma_R \left( \begin{array}{cccc} \cos^2\theta &  \cos\theta \sin\theta \\ \cos\theta \sin\theta &
   \sin^2\theta  \end{array} \right).
\end{aligned}
\end{equation}

To break the left-right {reflection} symmetry and {to} induce efficient energy filtering, {we assume that} the tunnel coupling to be Lorentzian {functions, following Ref.~\onlinecite{Rafael},}
\begin{equation}
\begin{aligned}
\Gamma_L = \Gamma_0\frac{\Gamma_0^2}{(\omega-E_l)^2+\Gamma_0^2}, \ \Gamma_R = \Gamma_0\frac{\Gamma_0^2}{(\omega-E_r)^2+\Gamma_0^2}.
\end{aligned}
\end{equation}
After obtaining the c-DQD Green's functions (see Appendix \ref{green}) the elastic and inelastic {electric} currents {flowing into the $L$ lead} can be calculated (see Appendix \ref{currapp}) as ($e<0$ is the electronic charge)
\begin{subequations}
\begin{align}
& I_e^L|_{el} = e\int \frac{d\omega}{2\pi} {\rm Tr}(\hat \Gamma^L_{rot}(\omega)\hat G_{tot}^r(\omega)[\hat \Sigma_l^<(\omega) \nonumber \\
& \hspace{1cm} + 2f_L(\omega)\hat \Sigma_l^r(\omega)]\hat G_{tot}^a(\omega)) ,\\
& I_e^L|_{inel} =  e\int \frac{d\omega}{2\pi} {\rm Tr}(\hat \Gamma^L_{rot}(\omega)\hat G_{1}^r(\omega)[\hat \Sigma_P^<(\omega) \nonumber \\
& \hspace{1cm} + 2f_L(\omega)\hat \Sigma_P^r(\omega)]\hat G_{1}^a(\omega)).
\end{align}
\end{subequations}
The elastic and inelastic heat currents flowing into the $L$ lead is calculated as
\begin{subequations}
\begin{align}
& I_Q^L|_{el} = \int \frac{d\omega}{2\pi} (\omega-\mu_L){\rm Tr}(\hat \Gamma^L_{rot}(\omega)\hat G_{tot}^r(\omega)[\hat \Sigma_l^<(\omega)\nonumber \\
& \hspace{1cm} + 2f_L(\omega)\hat \Sigma_l^r(\omega)]\hat G_{tot}^a(\omega)), \\
& I_Q^L|_{inel} = \int \frac{d\omega}{2\pi} (\omega-\mu_L){\rm Tr}(\hat \Gamma^L_{rot}(\omega)\hat G_{1}^r(\omega)[\hat \Sigma_P^<(\omega)\nonumber \\
&\hspace{1.2cm} + 2f_L(\omega)\hat \Sigma_P^r(\omega)]\hat G_{1}^a(\omega)).
\end{align}
\end{subequations}
where $f_L=[e^{(\omega-\mu_L)/k_BT_L}+1]^{-1}$ is the Fermi-Dirac distributions for $L$ reservoir. For the charge and heat currents flowing into the $R$ lead, the same expressions hold once $L\rightarrow R$. Here, $\mu_{L(R)}=\mu\pm\Delta\mu/2$, $\mu$ is the equilibrium chemical potential and $\Delta\mu$ is the electrochemical potential bias.  Charge conservation implies that $I_e^L+I_e^R=0$, while energy conservation requires $I_Q^L+I_Q^R+I_Q^P+\mu_L I_e^L/e+\mu_R I_e^R/e=0$. The net charge current flowing from the left reservoir to the right one is then
\begin{align}
I_e=\frac{1}{2}(I_e^R-I_e^L) .
\end{align}
The heat current flowing into the photonic bath is
\begin{align}
I_Q^P=-(I_Q^L+I_Q^R+\frac{\mu_L}{e} I_e^L+ \frac{\mu_R}{e} I_e^R ) ,
\end{align}
and the net heat current exchanged between the $L$ and $R$ leads (from $L$ to $R$) is
\be
I_Q =\frac{1}{2} (I_Q^R-I_Q^L) .
\ee

{Before going into the details, we briefly state the methods and approximations used in this work. It was found in
Ref.~\onlinecite{RenPRB} that, in the limit of vanishing dot-lead coupling, the exact Green's function for electrons in the quantum-dots 
can be obtained using the polaron eigenstates for {\em arbitrary strong} light-matter interaction. In the regime with finite but very small 
dot-lead coupling (smaller than any other energy scale of the system), one can treat the dot-lead coupling using perturbation theory. 
Up to the lowest order perturbation (i.e., the linear order of the dot-lead tunneling rates), the transport currents can be formulated 
using the Green's function approach. The details of the polaron eigenstates, the Green's functions and the Feynman diagrams are 
given in Appendix A. The transport currents and the Feynman diagrams for higher-order corrections are given in the Appendix B. 
The dot-lead coupling introduces both elastic and inelastic transport effects. The former has been treated in a non-perturbative way in 
Ref.~\onlinecite{RenPRB} by one of the authors. The latter is treated for the first time in this work using the polaron Green?s 
function method. Specifically, we treat the inelastic transport currents using the lowest order perturbation in the dot-lead coupling, 
which is justified for weak dot-lead coupling. The light-matter interaction is treated in the non-crossing approximation beyond the $g^2$ 
order. The crossing Feynman diagrams ignored in this work are of $g^4$ and beyond.}

{Because of the non-crossing approximation, our theory is valid in the region $g<0.1$ (i.e., $g^2<0.01$) where higher-order corrections 
are negligible. In some of the figures (Figs. 1 and 3) we extend the calculations to $0.1<g<0.25$, to see the qualitative trends beyond 
the region of $g<0.1$. We believe that these qualitative trends are meaningful, since $g^2<0.0625$ and the next order correction is still fairly small.} In Fig.~\ref{fig1}(d), we show the dependences of the elastic and inelastic electric currents on the electron-photon interaction which is characterized by the dimensionless parameter $g$. Compared with the inelastic electric current, the elastic electric current has much weaker dependences on the light-matter interaction, since it does not rely crucially on the light-matter coupling. However, the light-matter interaction does modify the elastic electric current, mainly due to the following two mechanisms: the shift of the electronic energy due to the polaron effect, i.e., $E_\alpha\rightarrow E_\alpha-g^2\omega_c$ ($\alpha=D,d$), and the side-bands effect. These two effects are sensitive to the chemical potential which determines the distribution on the main peak and the side-bands. Therefore, the chemical potential can significantly
change the dependence of the elastic electric current on the light-matter interaction, as shown in Fig.~1(d). The dependence of the inelastic electric current on the light-matter interaction is distinct from that of the elastic electric current. {From the figure, we find that the inelastic electric current is proportional to $g^2$ for small $g$ (i.e., weak light-matter interaction), which is consistent with the rate equation 
results \cite{Jiang2012}. The dependence becomes much stronger, starting at $g\simeq 0.03$, where our perturbation approach is still
valid. This result indicates that, although our treatment of the inelastic currents is perturbative, it goes beyond the conventional
rate equation approach (which always predict $\sim g^2$ dependence for the currents). This is because we used the polaron
Green's functions which contain higher-order corrections due to light-matter interaction, beyond the bare electron Green's function.
We remark that the change of the power-law dependence in the inelastic currents is one of the featuring results in this work due to
strong light-matter interaction.}

\begin{figure}[htb]
  \centering \includegraphics[width=4.2cm]{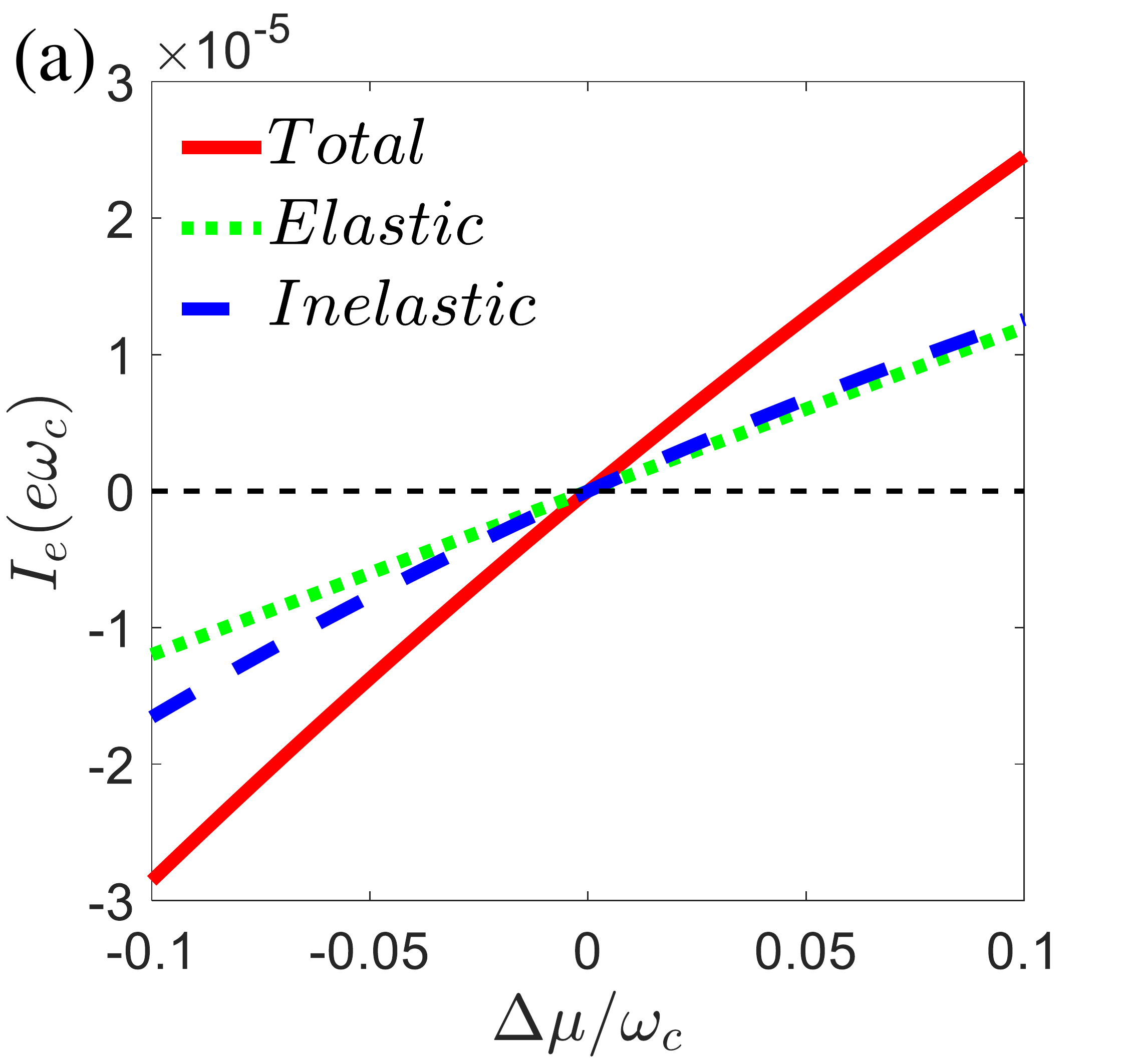}\hspace{0.2cm}\includegraphics[width=4.2cm]{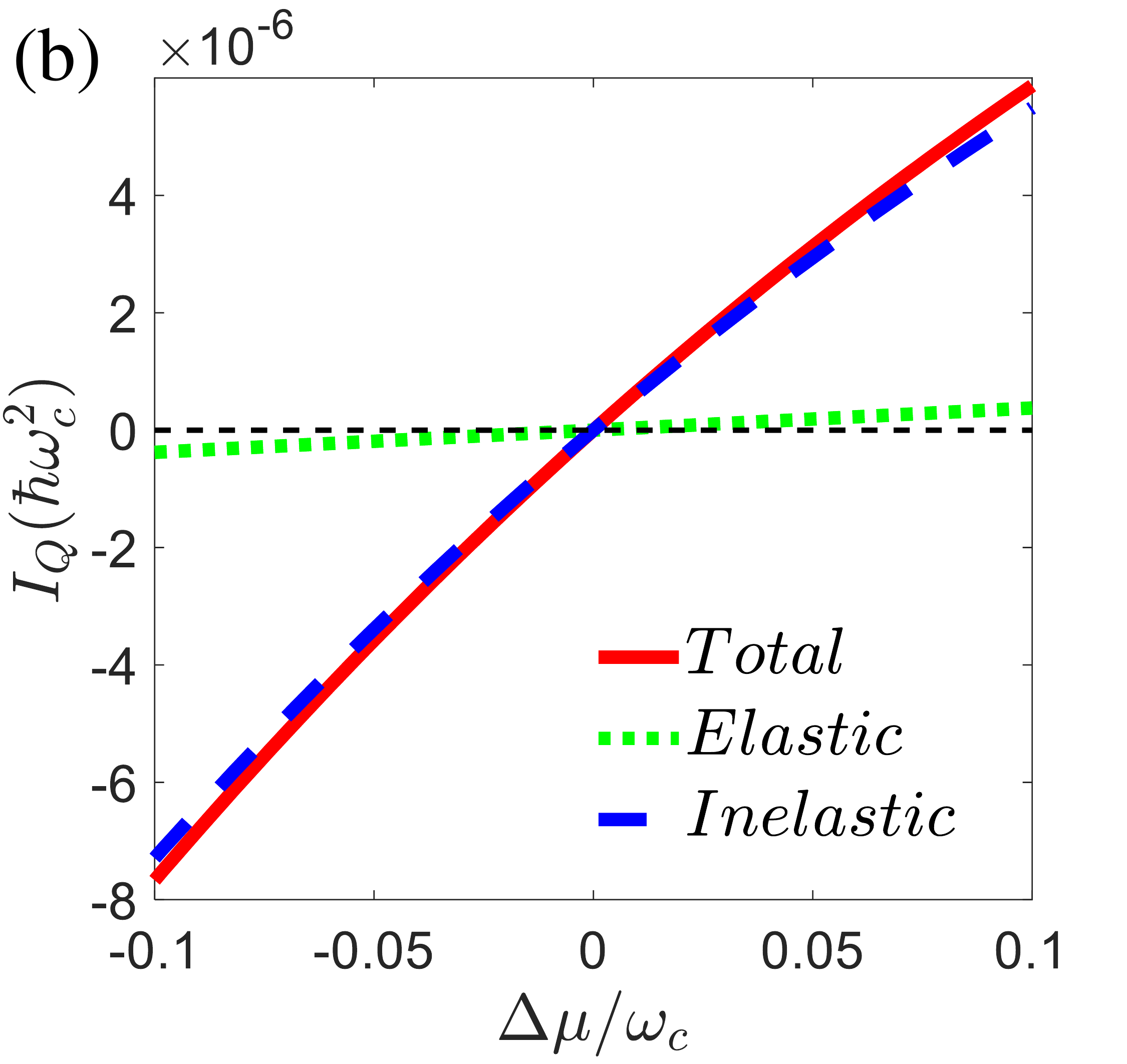}
  \centering \includegraphics[width=4.2cm]{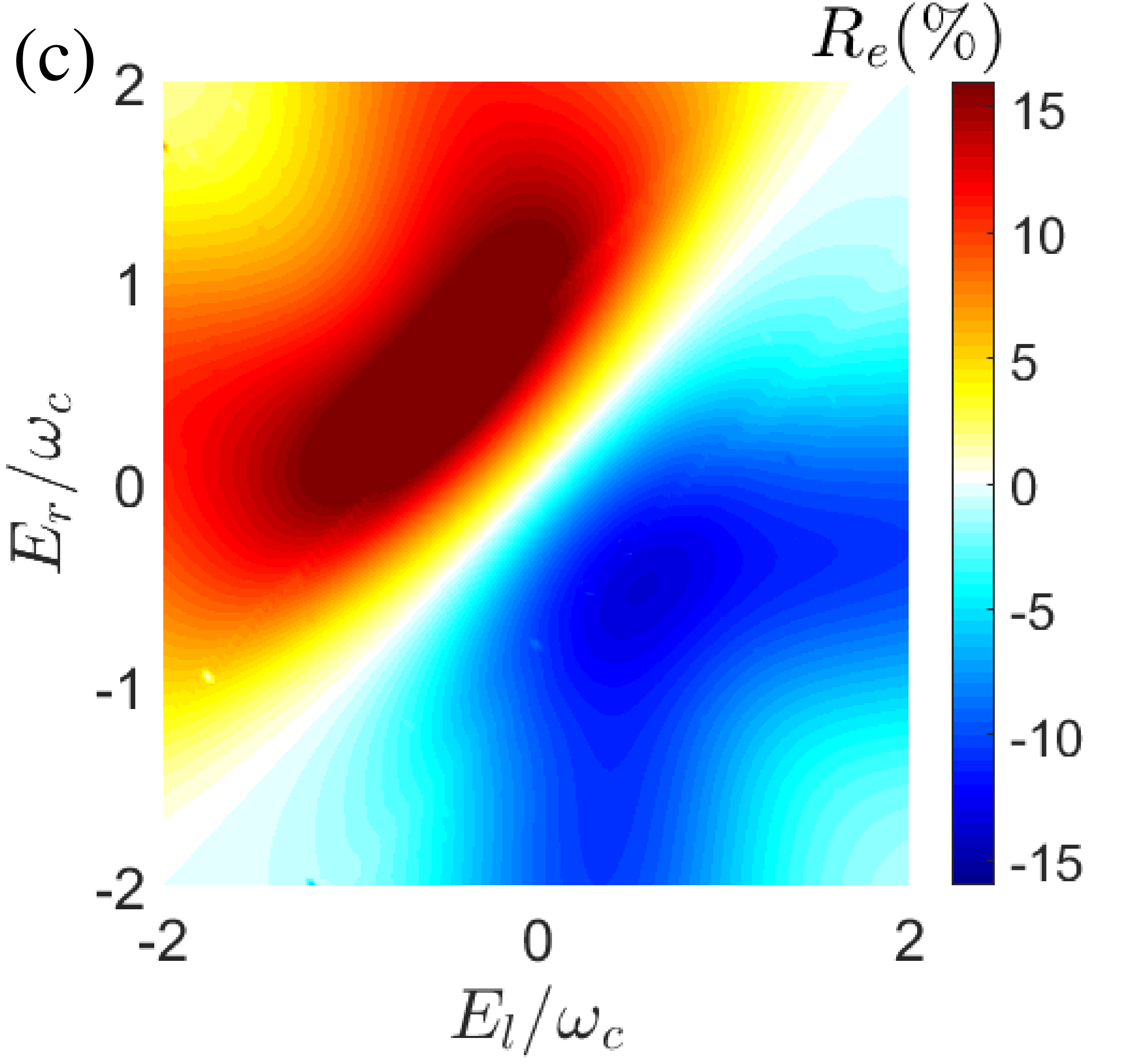}\hspace{0.2cm}\includegraphics[width=4.2cm]{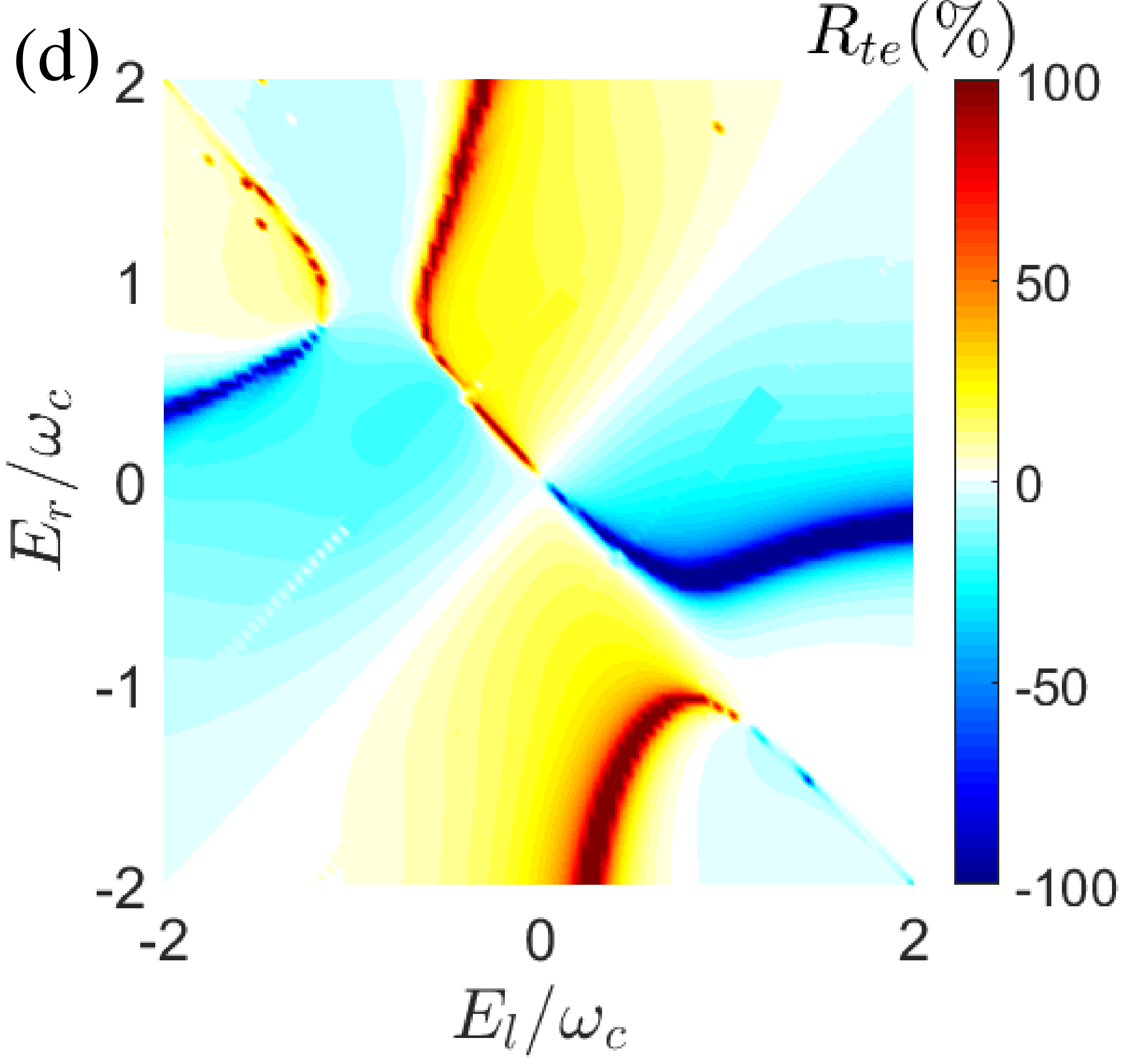}
 \caption{(Color online)  The current (a) and the heat current (b) as the function of $\Delta\mu$ with $\mu=0$. The parameters are $k_BT=0.1\omega_c$, $E_l=0$, $E_r=2.0\omega_c$, $t=0.3\omega_c$, $\Gamma_0=0.1\omega_c$ and $g=0.5$. (c) Charge rectification $R_e$ and (d) cross rectification $R_{te}$ as the function of $E_l$ and $E_r$ for $g=0.1$. The other parameters are $k_BT=0.2\omega_c$, $\mu=0$, $\Delta\mu=0.1\omega_c$ $t=0.3\omega_c$, $\Gamma_0=0.1\omega_c$.}
\label{fig2}
\end{figure}

\begin{figure}[htb]
  \centering \includegraphics[width=4.2cm]{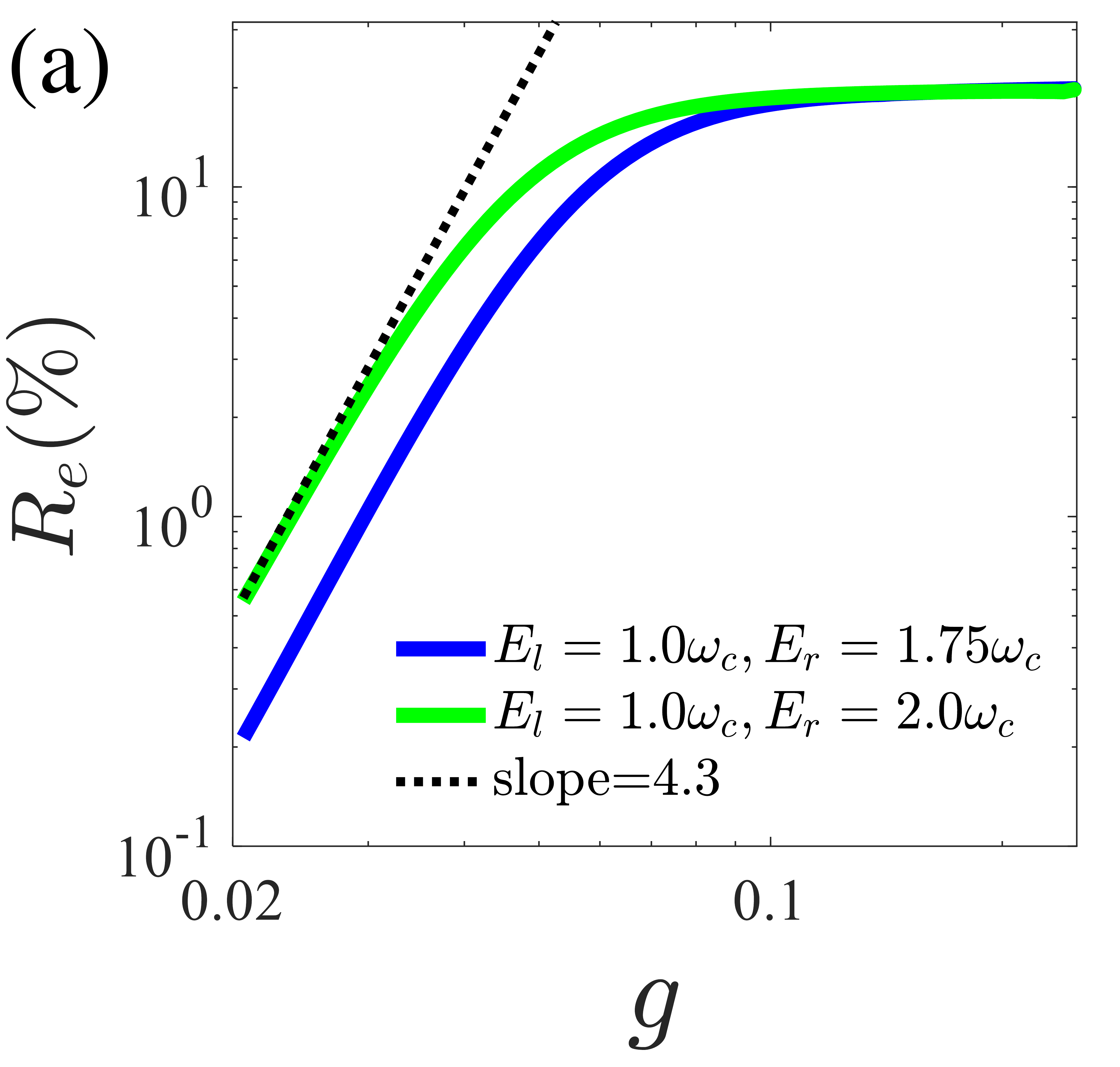}\hspace{0.2cm}\includegraphics[width=4.2cm]{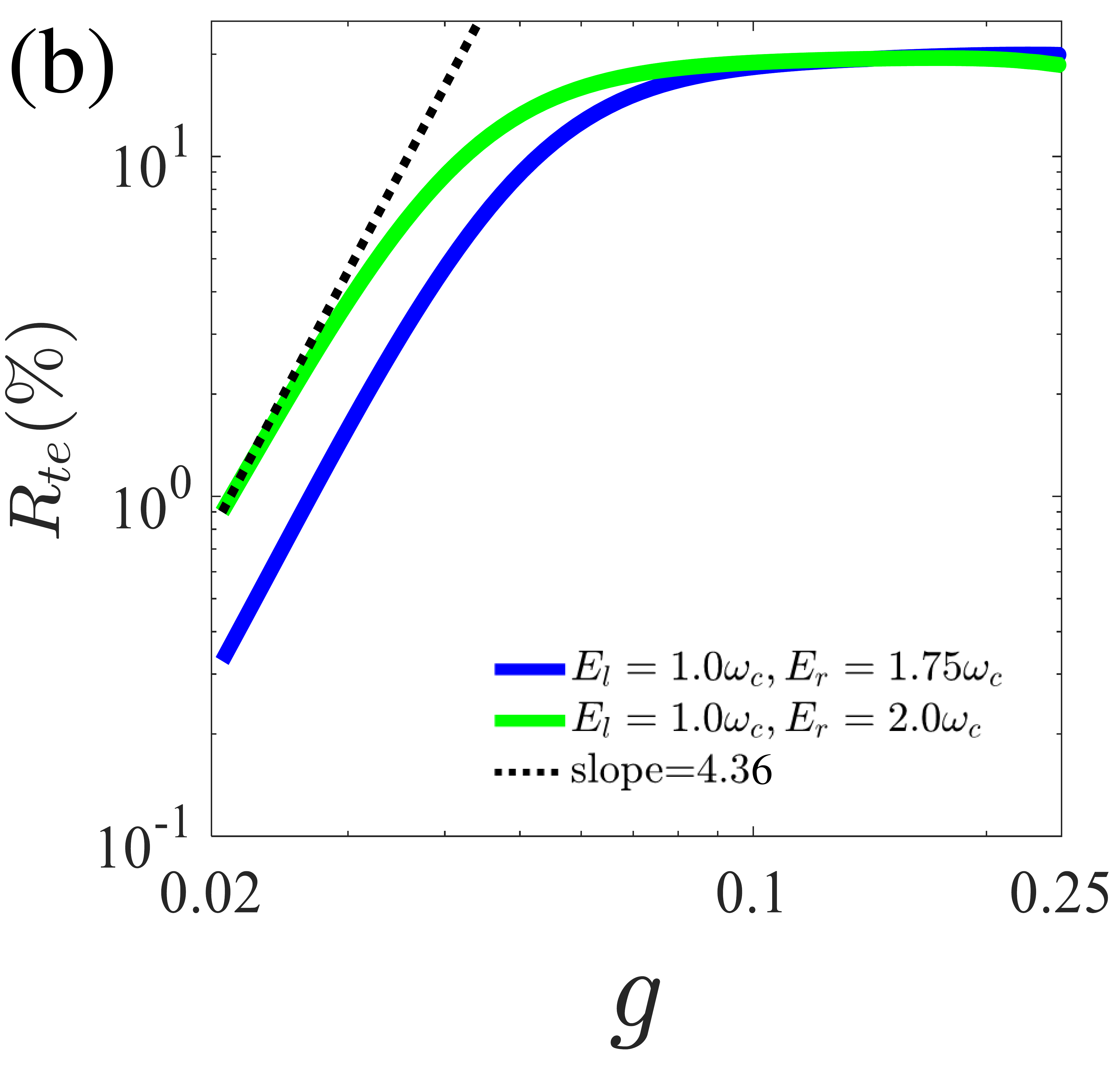}
\caption{(Color online) (a) Charge and (b) Peltier rectifications, $R_e$ and $R_{te}$, as functions of the light-mater interaction parameter $g$ for different $E_l$ and $E_r$ with $k_BT=\Delta\mu=0.1\omega_c$. Other parameters are $t=0.3\omega_c$, $\Gamma_0=0.01\omega_c$ and $\mu=0$ for all figures.}
\label{figB3}
\end{figure}

\section{Thermoelectric rectification effects}
{We now study the thermoelectric rectification effects. In thermoelectric systems, beside the conventional charge and heat rectification
effects, there are also cross-rectification effects between charge and heat~\cite{Jiangtransistors}. For instance, there can be charge rectification induced by temperature differences. This phenomenon can be denoted as Seebeck rectification, since it reflects the asymmetry of the Seebeck effect
with respect to the forward and backward temperature differences. Similarly, there can be heat rectification induced by voltage biases, which
measures the asymmetry of the Peltier effect with respect to forward and backward voltages and hence is denoted as Peltier rectification.
To the best of our knowledge, there is still no study on such cross-rectification effects in c-QED systems. The amplitude of the rectification effects is calibrated by}
\begin{equation}
\begin{aligned}
R_e=\frac{I_e(\Delta\mu)+I_e(-\Delta\mu)}{|I_e(\Delta\mu)|+| I_e(-\Delta\mu)|},
\end{aligned}
\end{equation}
for the charge rectification, and
\begin{equation}
\begin{aligned}
R_{te}=\frac{I_Q(\Delta\mu)+I_Q(-\Delta\mu)}{|I_Q(\Delta\mu)|+| I_Q(-\Delta\mu)|},
\end{aligned}
\end{equation}
for the Peltier rectification. Typical electrochemical potential differences $\Delta\mu$ for pronounced rectification effects 
are comparable with $k_BT$.

In Fig.~\ref{fig2}, we demonstrate and study the charge and Peltier rectifications. The asymmetric charge and heat transport with respect to the forward and backward voltage biases are shown in Figs.~\ref{fig2}(a) and \ref{fig2}(b). We find that the asymmetry is induced by the inelastic transport processes. As shown in Ref.~\onlinecite{RenPRB}, the elastic currents are anti-symmetric with respect to forward and backward voltage and temperature biases. Since the asymmetry only arises from the inelastic transport, the light-matter interaction plays the essential role for both charge and Peltier rectifications.
Strong light-matter interaction leads to strong rectification effects. Figs.~\ref{fig2}(c)-\ref{fig2}(d) give the dependences of the rectification effects on the QDs energies. There are hot-spots for both charge and Peltier rectifications. For instance,  charge rectification is 
pronounced at certain energies where $E_\ell$ is considerably different from $E_r$. Peltier rectification is more sensitive to the QDs 
energies. Both the charge and Peltier rectification coefficients are anti-symmetric around the line of $E_\ell=E_r$. However, these results are considerably different from the weak coupling regime where the Peltier rectification is also anti-symmetric with respect to the line of
$E_\ell=-E_r$~\cite{Jiangtransistors}. We understand that this is mainly due to the polaron-induced energy shift, i.e., $E_\alpha\rightarrow E_\alpha-g^2\omega_c$ ($\alpha=D,d$), and the side band effects.

Fig.~\ref{figB3} shows the dependences of the charge and Peltier rectifications on the light-matter interaction for two different QDs energies. General trends can be observed {from the figure: for small $g$ (i.e., weak light-matter interaction), the dependences follow a power-law $\sim g^{\gamma}$ with $\gamma\sim 4$ but depends on specific QDs energies, temperatures and electrochemical potential differences; for large $g$ (i.e., strong light-matter interaction), the power-law dependences are not valid any more. Since the linear transport coefficients due to inelastic transport processes are proportional to $g^2$ for small $g$, the rectification coefficients, which is due to nonlinear transport effects, should be proportional to higher exponents. The observed power-law dependences with exponents $\gamma\sim 4$ agree with such arguments. The power-law dependences indicate that pronounced rectification effects require generally strong light-matter interaction.}

\section{Thermal transistor effect in the linear transport regime}
{It was well accepted for a long time that nonlinear transport is the prerequisite for thermal transistor effects. In particular, negative differential
thermal conductance is believed to be the necessary condition for the emergence of thermal transistor effects~\cite{RenRMP}. It was first argued in Ref.~\onlinecite{Jiangtransistors} that thermal transistor effects can emerge in the linear-transport regime if phonon-assisted inelastic transport is dominant. However, the rate equation method used in Ref.~\onlinecite{Jiangtransistors} is valid only when the electron-phonon interaction is very weak. Here we show, using the more rigorous Green's function method, that such {\em linear} thermal transistor effect also exists in c-QED systems for a large range of QDs energies and light-matter interactions}. If we consider purely thermal conduction (i.e., the electrochemical potential difference is set to zero), the linear thermal transport properties of the system can be described by\cite{Transistor1,Transistor2,transistor3}:
\begin{equation}
\begin{aligned}
\left( \begin{array}{cccc} I_Q^P\\ I_Q^R \end{array}\right) =
 \left( \begin{array}{cccc} K_{PP} & K_{PR} \\ K_{RP} &
    K_{RR} \end{array} \right) \left( \begin{array}{cccc} T_P-T_L\\
    T_R-T_L \end{array}\right),
\end{aligned}
\end{equation}
where $K_{PP}=\frac{\partial I_Q^P}{\partial T_P}$, $K_{PR}=\frac{\partial I_Q^P}{\partial T_R}$, $K_{RP}=\frac{\partial I_Q^R}{\partial T_P}$ and $K_{RR}=\frac{\partial I_Q^R}{\partial T_R}$ in the limit $T_L, T_R, T_P\rightarrow T$.
From the above, the heat current amplification factor is given by
\begin{equation}
\begin{aligned}
\alpha=\Big|\frac{\partial_{T_P}I_Q^R}{\partial_{T_P}I_Q^P}\Big|=\frac{K_{RP}}{K_{PP}},
\end{aligned}
\end{equation}

\begin{figure}[htb]
  \centering \includegraphics[width=4.2cm]{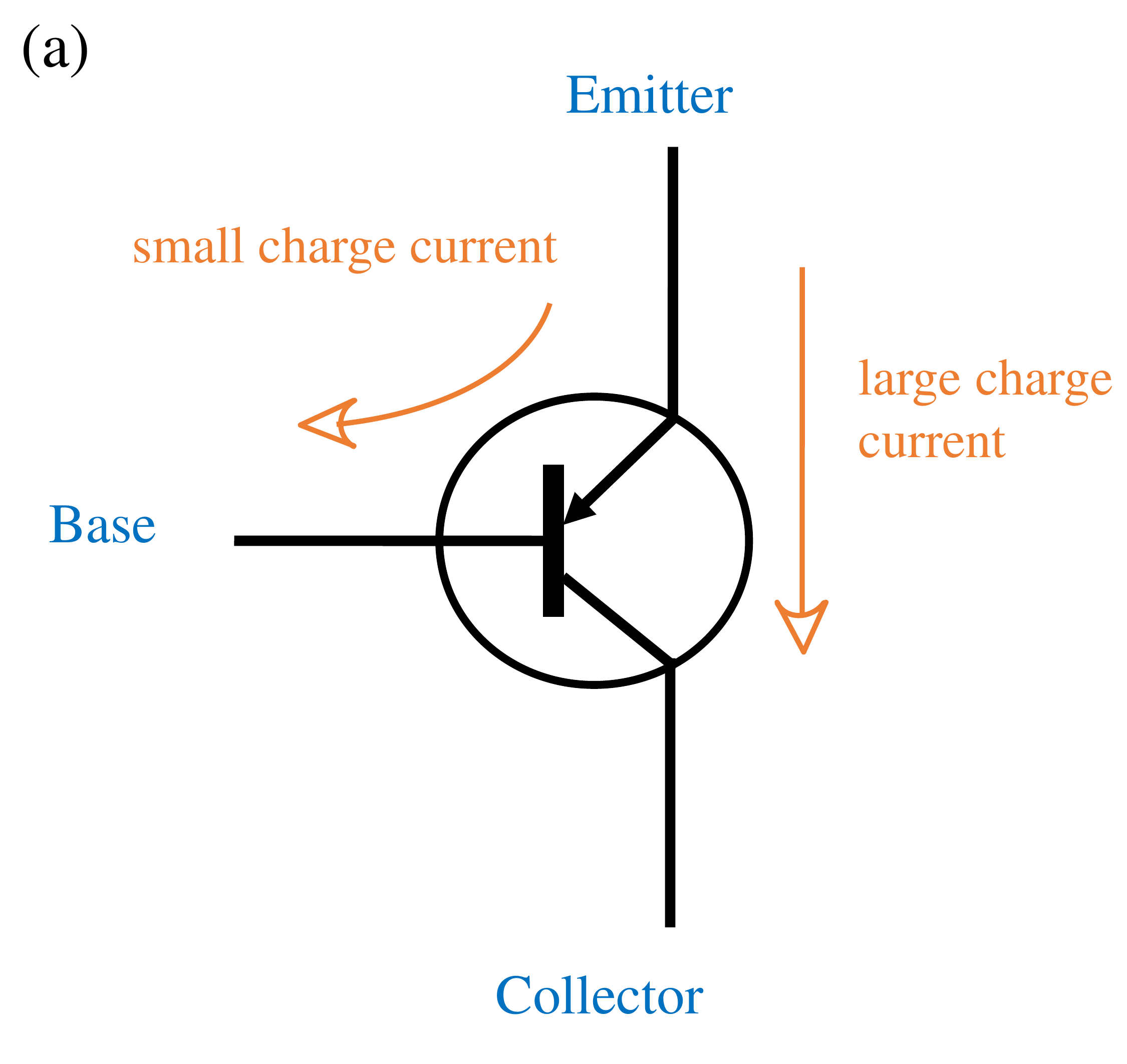}\hspace{0.2cm}\includegraphics[width=4.2cm]{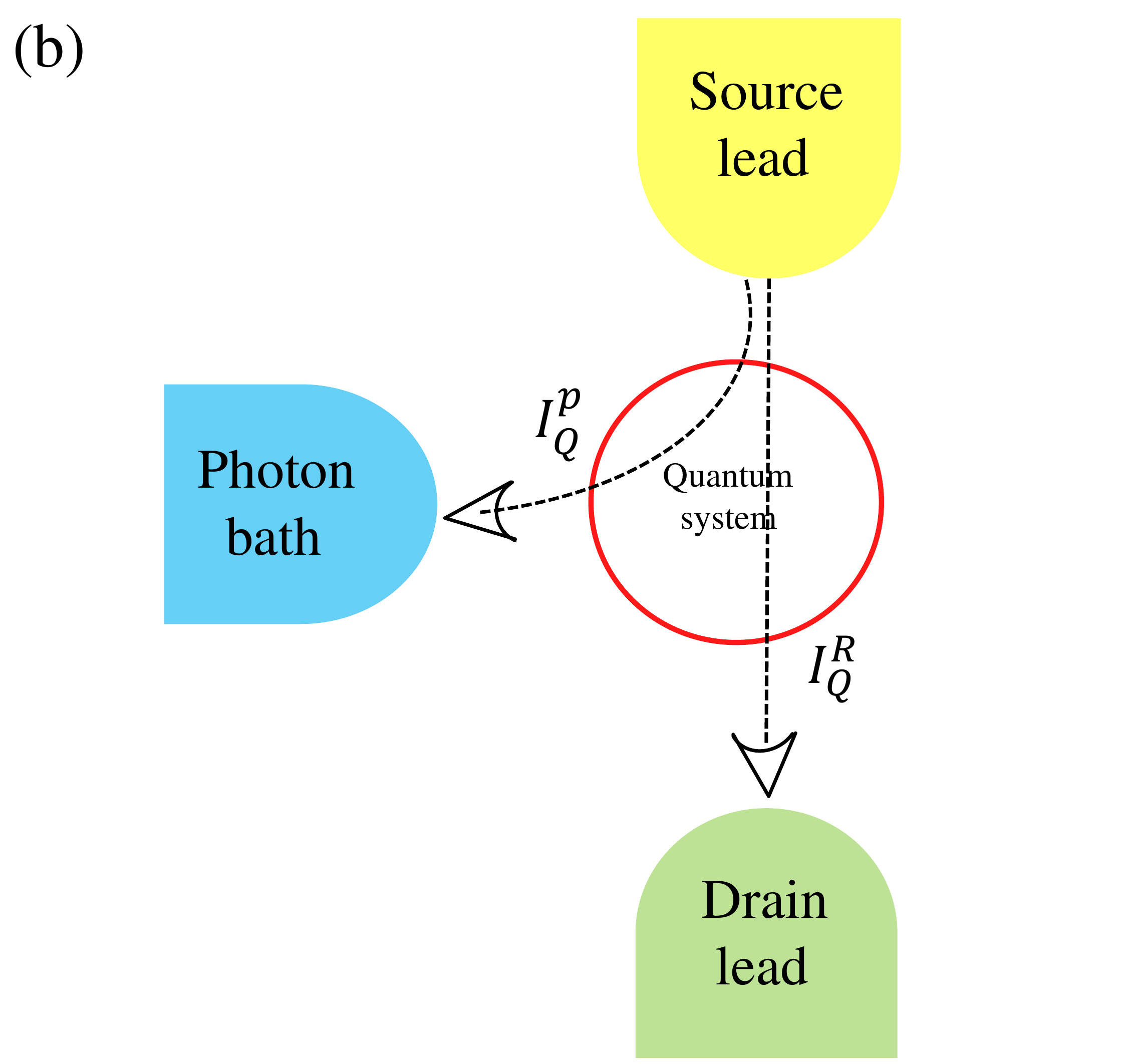}
  \centering \includegraphics[width=4.2cm]{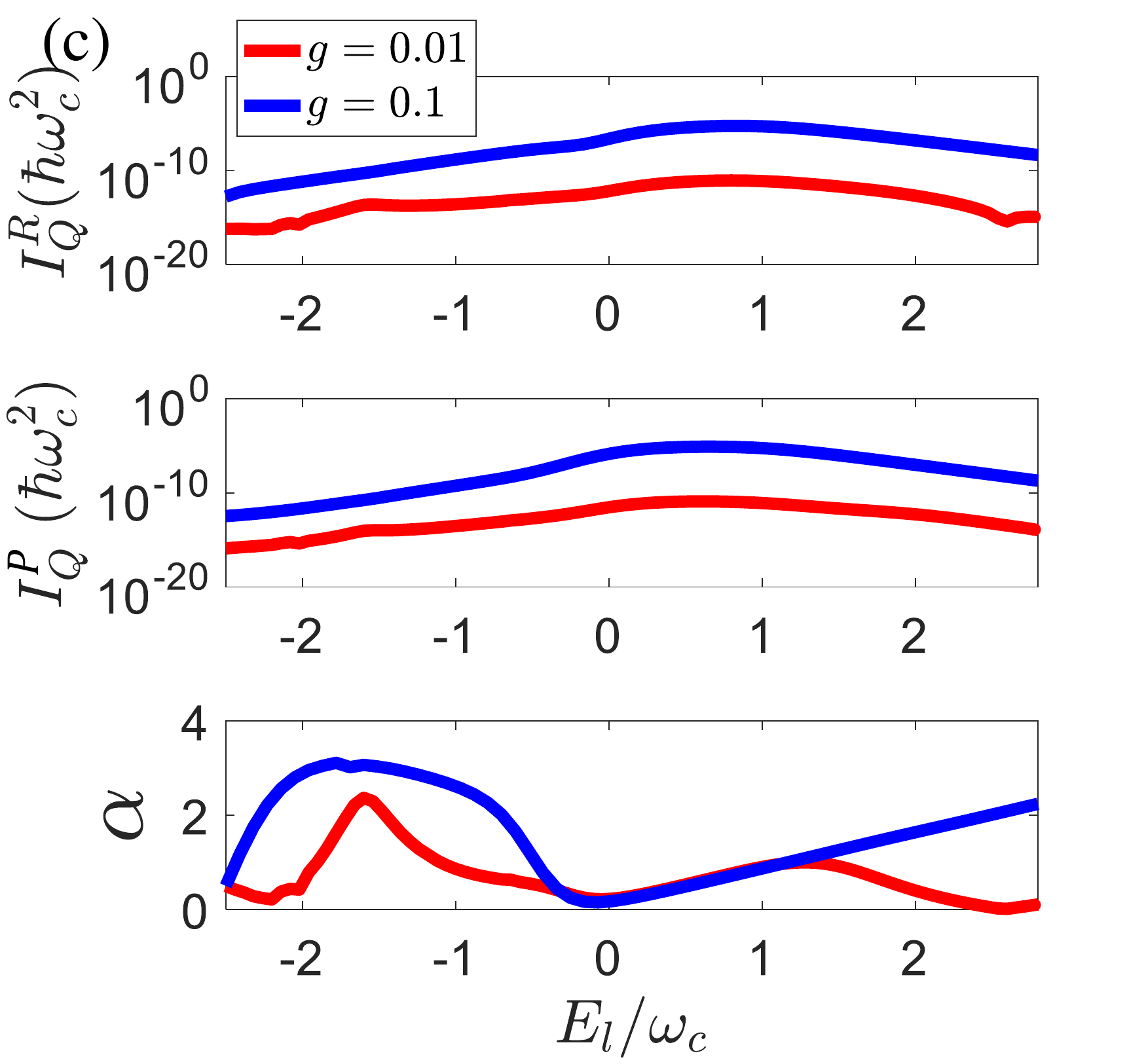}\hspace{0.2cm}\includegraphics[width=4.2cm]{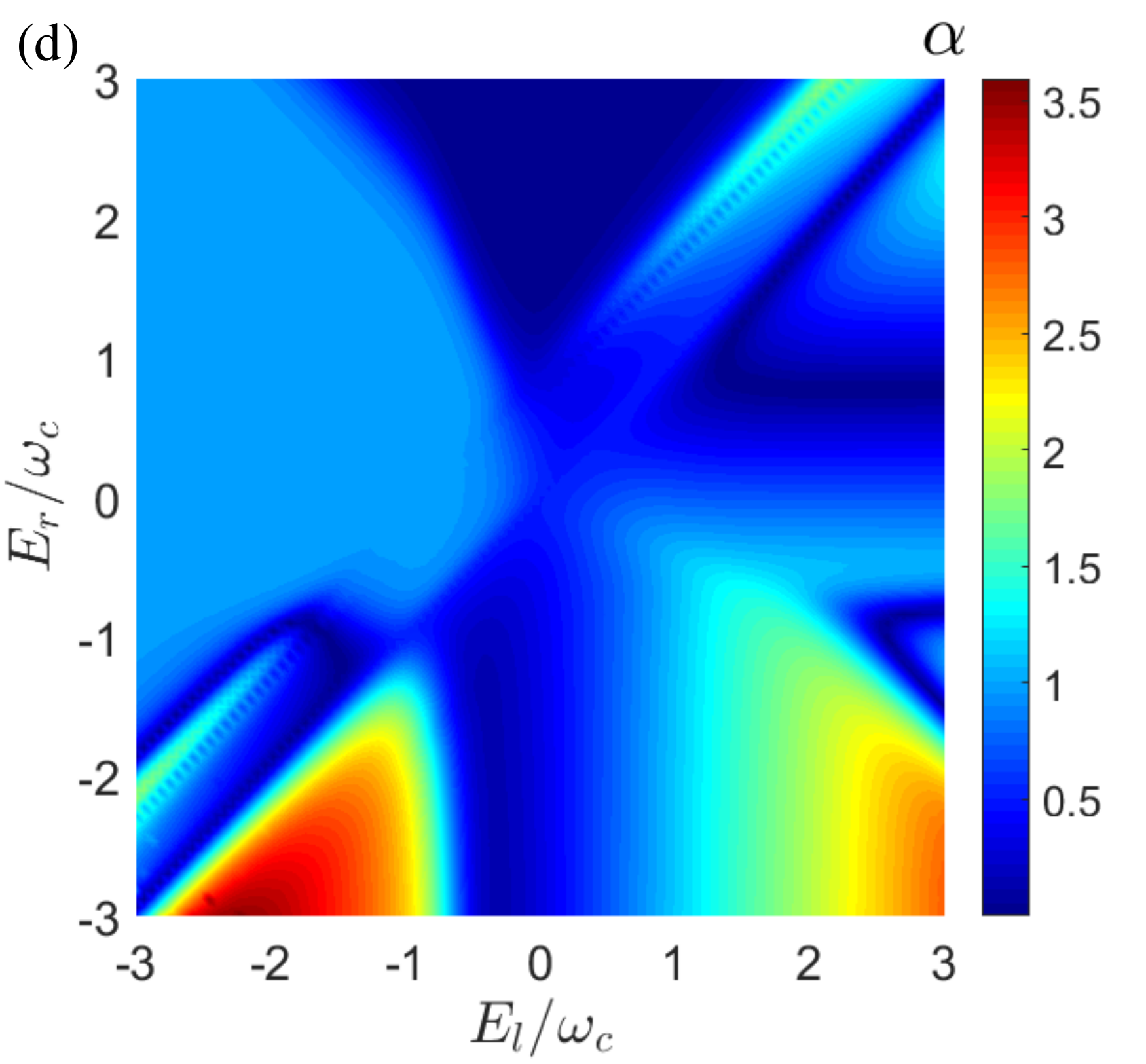}
 \caption{(Color online) (a) Schematic of the function of a conventional transistor. A small charge current flowing from the emitter to the base controls a large charge current flowing from the emitter to the collector. The ratio between the two charge currents is the quantity that characterizes the transistor effect. (b) Schematic of the DQD---c-QED system as a thermal transistor. The DQD---c-QED system acts like the PNP junction. The small heat current flowing from the source lead to the photon bath, $I_Q^P$, can controle the large heat current flowing from the source lead to the drain lead, $I_Q^R$. The ratio between the two heat currents defines the heat current amplification factor, $\alpha$, which characterizes the thermal transistor effect. (c) The heat current flowing into the drain lead, $I_Q^R$, the photonic heat current $I_Q^P$, and the heat current amplification factor $\alpha$, as functions of the QD energy $E_\ell$ for two different light-matter interactions $g=0.01$ and $0.1$, where the other parameters are $k_BT=0.2\ome_c$, $\mu=1.0\ome_c$, $E_r=-2.5\ome_c$ and $\Gamma_0=0.1\ome_c$. The heat currents and heat current amplification factor $\alpha$ are considerably larger for the strong light-matter interaction. (d) The heat current amplification factor $\alpha$ as a function of the QDs energies $E_\ell$ and $E_r$. Here $g=0.1$ while ther parameters are the same as in figure (c).}\label{fig3}
\end{figure}

As schematically illustrated in Figs.~\ref{fig3}(a) and \ref{fig3}(b), the condition for thermal transistor is $\alpha>1$ \cite{Jiangtransistors}. In Figs.~\ref{fig3}(c) and \ref{fig3}(d), we find that the coefficient $\alpha$ is very sensitive to the
QDs energies which can be controlled easily via gate-voltages in experiments. In particular, there are hot-spots for $\alpha$ to be considerably larger than 1, {particularly for $1<|E_\ell/\ome_c|<3$, while $-3<E_r/\ome_c<-1$. Detailed dependences of the heat currents and the thermal transistor coefficient $\alpha$ on the QD energy $E_\ell$ is shown in Fig.~\ref{fig3}(c) when $E_r=-2.5\ome_c$.} It is shown that pronounced thermal transistor effect can be achieved at considerably large heat currents
when the light-matter interaction is strong. In general, strong light-matter interaction helps the thermal transistor effect  in the linear-transport regime. It also enhances the heat currents significantly, {since the photon heat current is proportional to the inelastic transition rate which increases rapidly with the light-matter interaction as shown in Fig.~1.}

\section{Conclusion and outlook} 
We have shown that QD systems placed at finite voltage bias and integrated with a superconducting c-QED architecture can serve as excellent charge and Peltier rectifiers (diodes). Thermal transistor effects in the linear transport regime is also found thanks to photon-assisted inelastic transport. Although the paper is primarily discussed for a QD c-QED architecture, our results are very applicable to molecular  junctions\cite{OraPRB2010,Fluctuationtheorem2,RenPRB,simine2012vibrational,simine2013path,xu2015negative,simine2015can,Agarwalla2015,Dvira} as well where the role of photons is played by the molecular vibrations. {However, we have not considered the role of 
electron-phonon interactions and electron-electron Coulomb interactions, which generally exist in QDs c-QED systems\cite{PhysRevLett.117.056801}. Our results are applicable when electron-phonon interaction is much weaker than the light-matter interaction, the intra-dot Coulomb interaction is very strong, while the inter-dot Coulomb interaction is negligible}. Future work will involve understanding the role of electron-phonon interactions and studying the impact of onsite and inter-site Coulumb interactions\cite{Kontos,Delattre,kontos_kondo}, {alongside with the fundamental strong light-matter interaction effects\cite{PhysRevX.4.031025}}.

\section*{Acknowledgments}
J.L., R.W., and J.-H.J. acknowledge support from the National Natural Science Foundation of China (NSFC Grant No. 11675116), Jiangsu Distinguished Professor funding, and a Project Funded by the Priority Academic Program Development of Jiangsu Higher Education Institutions (PAPD). M.K. gratefully acknowledges the Ramanujan Fellowship No.SB/S2/RJN-114/2016 from the Science and Engineering Research Board (SERB), Department of Science and Technology, Government of India. M. K. gratefully acknowledges useful discussions with Jason Petta, Takis Kontos, Aashish Clerk and Aditi Mitra. J.R. acknowledges support from the NNSFC with Grant No.11775159, the Natural Science Foundation of Shanghai (No. 18ZR1442800), and the National Youth 1000 Talents Program in China. We acknowledge support from the International Centre for Theoretical Sciences, Tata Institute of Fundamental Research during the program on 'open quantum systems', ICTS/Prog -oqs2017/2017/07. J. L. acknowledges the hospitality of ICTS-TIFR.

\appendix
\section{Non-perturbative Green's functions for the DQD c-QED model without dot-lead coupling}
\label{green}
We start by analytically solving the eigenproblem for the DQD c-QED model. Following the relation $G^r(t)=\Theta(t)(G^>(t)-G^<(t))$,
$G^a(t)=-\Theta(-t)(G^>(t)-G^<(t))$, and utilizing $\Theta(t)=\int
\frac{d\omega}{2\pi i}\frac{e^{i\omega t}}{\omega-i0^{+}},$ we have
the retarded (advanced) Green's function:
\begin{eqnarray}
G_{0D}^{r/a}(\omega) =&
\int\frac{d\omega_1}{2\pi}\int\frac{d\omega_2}{2\pi i}\int dt
e^{i\omega t}\frac{e^{-i(\omega_1-\omega_2)t}}{\omega_2\mp i0^{+}}\nonumber \\
&\times[G_{0D}^>(\omega_1)-G_{0D}^<(\omega_1)].
~\label{retarded}
\end{eqnarray}

Following the method of Ref.~\onlinecite{RenPRB}, we first detail the calculation of the lesser Green's function
\begin{align}
G_{0D}^<(\omega)=& i\int^{+\infty}_{-\infty}dt e^{i\omega
t}\langle d_{D}^{\dag}(0)d_{D}(t)\rangle   \nonumber\\
=&i\int^{+\infty}_{-\infty}dt
e^{i\omega t}\sum_{\varphi}\sum_{\psi}\langle\varphi|\rho
d_{D}^{\dag}(0)|\psi\rangle    \nonumber\\ &\times\langle\psi|e^{iH_{c-DQD}t}d_{D}(0)e^{-iH_{c-DQD}t}|\varphi\rangle,
\end{align}
where $\rho=e^{-\beta H_{c-DQD}}/Z$ with $Z=\text{Tr}(e^{-\beta H_{c-DQD}})$. Here $|\varphi\rangle$ and $|\psi\rangle$ are the possible eigenstates.

We introduce a cavity photon basis with displacements shifted by different QD states through the e-p coupling\cite{RenPRB}
\begin{equation}
|n\rangle_{\nu}=[{(\hat{A}^{\dag}_{\nu})^n}/{\sqrt{n!}}]
\exp{(-g_{\nu}^2/2-g_{\nu}\hat{a}^{\dag})}|0\rangle,
\end{equation}
where $\hat{A}^{\dag}_{\nu}=\hat{a}^{\dag}+g_{\nu}$ denotes the creator that creates a photon displaced from the original position by a value $g_{\nu}$ depending on the electronic state, that is, $g_{0}=0$, $g_{D}=g_{d}=g$, and
$g_{Dd}=g_{D}+g_{d}=2g$, $n=0,1,2...$.
Therefore, with the help of the cavity photon basis, the solution to the eigenvalue problem is
\begin{eqnarray}
{_{0}\langle 0,n}|H_{c-DQD}|0,n\rangle_{0}&=&n\omega_c,\\
{_{D}\langle D,n}|H_{c-DQD}|D,n\rangle_{D}&=&n\omega_c+\tilde E_{D},\\ ~\label{shift}
{_{d}\langle d,n}|H_{c-DQD}|d,n\rangle_{d}&=&n\omega_c+\tilde E_{d},\\
{_{Dd}\langle Dd,n}|H_{c-DQD}|Dd,n\rangle_{Dd}&=&n\omega_c+\tilde E_{Dd},
\end{eqnarray}
where $\tilde E_{D}=E_{D}-\omega_cg_D^2$, $\tilde E_{d}=E_{d}-\omega_cg_d^2$ and $\tilde E_{Dd}=\tilde E_{D}+\tilde E_{d}-2\omega_cg_Dg_d$.
Obviously, $|0, n\rangle_{0}$, $|D,n\rangle_{D}$, $|d,n\rangle_{d}$, $|Dd,n\rangle_{Dd}$ are four possible eigenstates and $n\omega_0$, $n\omega_c+\tilde E_{D}$, $n\omega_c+\tilde E_{d}$, $n\omega_c+\tilde E_{Dd}$ are the corresponding possible eigenvalues.

There are only two nonzero combinations for calculating $G_{D}^<(\omega)$: $|D,n\rangle_{D}$ and $|0,m\rangle_{0}$, or $|Dd,n\rangle_{Dd}$ and $|d,m\rangle_{d}$, and we
\begin{align}
G_{0D}^<(\omega) =& \frac{2\pi
i}{Z}\sum^{\infty}_{n=0}\sum^{\infty}_{m=0}\big[\delta(\omega-(n-m)\omega_c-\tilde E_{D}) \nonumber\\
&\times e^{-\beta
(n\omega_c+\tilde E_{D})}{_D\langle}
n|m\rangle_{0}\;{_{0}\langle} m|n\rangle_{D}
\nonumber\\
&+\delta(\omega-(n-m)\omega_c-(\tilde E_{Dd}-\tilde E_{d})) \nonumber\\
&\times e^{-\beta
(n\omega_c+\tilde E_{d})}{_{Dd}\langle}
n|m\rangle_{d}\;{_{d}\langle}
m|n\rangle_{Dd}\big].
\end{align}
The detailed expression of ${_b\langle} n|m\rangle_{c}$, denoting
the inner product of modified phonon states with effective
displacements $g_b$ and $g_c$, can be derived as follows:
\begin{eqnarray}
{_b\langle} n|m\rangle_{c} &=&
\langle0|\frac{(\hat{a}+g_b)^n}{\sqrt{n!}}
\exp{(-{g_b}^2/2-g_b\hat{a})} \nonumber\\
&&\times\frac{(\hat{a}^{\dag}+g_c)^m}{\sqrt{m!}}
\exp{(-{g_c}^2/2-g_c\hat{a}^{\dag})}|0\rangle \nonumber\\
&=& \frac{\exp{[-(g_b-g_c)^2/2]}}{\sqrt{n!
m!}}    \nonumber\\
&&\times\langle0|(\hat{a}+g_b)^n
e^{(-g_c\hat{a}^{\dag})}e^{(-g_b\hat{a})}(\hat{a}^{\dag}+g_c)^m
|0\rangle \nonumber\\
&=& \frac{\exp{[-(g_b-g_c)^2/2]}}{\sqrt{n!
m!}}    \nonumber\\
&&\times\langle0|(\hat{a}+g_b-g_c)^n (\hat{a}^{\dag}+g_c-g_b)^m
|0\rangle \nonumber\\
&=& \frac{\exp{[-(g_b-g_c)^2/2]}}{\sqrt{n!
m!}}    \nonumber\\
&&\times\sum^{\mathrm{min}\{n,m\}}_{k=0}k! C^{k}_{n}(g_b-g_c)^{n-k}C^{k}_{m}(g_c-g_b)^{m-k} \nonumber\\
&=& (-1)^m D_{nm}(g_b-g_c),
\end{eqnarray}
where
$$D_{nm}(x)=e^{-x^2/2}\sum^{\mathrm{min}\{n,m\}}_{k=0}\frac{(-1)^k\sqrt{n!
m!}x^{n+m-2k}}{(n-k)!(m-k)!k!}$$ is invariant under the exchange of
indices $n, m$. Note, to get the third equivalence, we utilized the
relation $\exp{(c\hat{a})}f(\hat{a}^{\dag},\hat{a})=f(\hat{a}^{\dag}+c,\hat{a})\exp{(c\hat{a})}$.

Therefore, the lesser Green's function can be further reduce to:
\begin{eqnarray}
G_{0D}^<(\omega)= \frac{2\pi
i}{Z}\sum^{\infty}_{n,m=0}\left[\delta(\omega-\Delta_{nm}^{(1)})e^{-\beta
(n\omega_0+\tilde E_{\sigma})}+\right. \nonumber\\
\left.\delta(\omega-\Delta_{nm}^{(2)})e^{-\beta
(n\omega_c+\tilde E_{Dd})}\right]D^2_{nm}(g_{D}),
\end{eqnarray}
where
\begin{align*}
\Delta_{nm}^{(1)} &=(n-m)\omega_c+\tilde E_{D}, \\
\Delta_{nm}^{(2)} &=(n-m)\omega_c+(\tilde E_{D}-2\omega_c g_{D}g_{d}), \\
D_{nm}(g_{D}) &=e^{-g_{D}^2/2}\sum^{\mathrm{min}\{n,m\}}_{k=0}\frac{(-1)^k\sqrt{n!m!}g_{D}^{n+m-2k}}{(n-k)!(m-k)!k!}, \\
Z &=(1+N_P)(1+e^{-\beta \tilde E_{D}}+e^{-\beta
\tilde E_{d}}+e^{-\beta\tilde E_{Dd}}).
\end{align*}
Here $N_P=1/(e^{\beta \omega_c}-1)$ denotes the Bose distribution of the photon population with inverse temperature
$\beta\equiv 1/k_BT_P$ and $\tilde E_{D}=E_{D}-\omega_cg_D^2$, $\tilde E_{d}=E_{d}-\omega_cg_d^2$ and $\tilde E_{Dd}=\tilde E_{D}+\tilde E_{d}-\omega_cg_Dg_d$.

Similarly, for the greater Green's function, we can obtain
\begin{eqnarray}
&G_{0D}^>(\omega)&=-i\int dt e^{i\omega t}\langle
d_{D}(t)d_{D}^{\dag}(0)\rangle  \nonumber\\
&=&\ -\frac{2\pi i}{Z}\sum_{n,m}\left[\delta(\omega-\Delta_{nm}^{(1)})e^{-\beta
m\omega_c}+\right.     \nonumber\\
&&\left.\delta(\omega-\Delta_{nm}^{(2)})e^{-\beta
(m\omega_c+\tilde E_{d})}\right]D^2_{nm}(g_{D})
\end{eqnarray}
Substituting the expressions of the greater and lesser Green'
functions into Eq.\ref{retarded}, we get the advanced and retarded Green's functions of the c-DQD
\begin{multline}
G_{0D}^{r/a}(\omega)=\frac{1}{Z}\sum^{\infty}_{n,m=0}\left[\frac{e^{-\beta
m\omega_c}+e^{-\beta(n\omega_c+\tilde E_{D})}}{\omega-\Delta_{mn}^{(1)}\pm
i0^{+}} + \right.\\
\left.\frac{e^{-\beta
(m\omega_c+\tilde E_{d})}+e^{-\beta
(n\omega_c+\tilde E_{d})}}{\omega-\Delta_{mn}^{(2)}\pm
i0^{+}}\right]D^2_{nm}(g_{D}),
\label{eq:G}
\end{multline}

\begin{figure}
\begin{center}
\includegraphics[width=3.00 in,height=3.00 in]{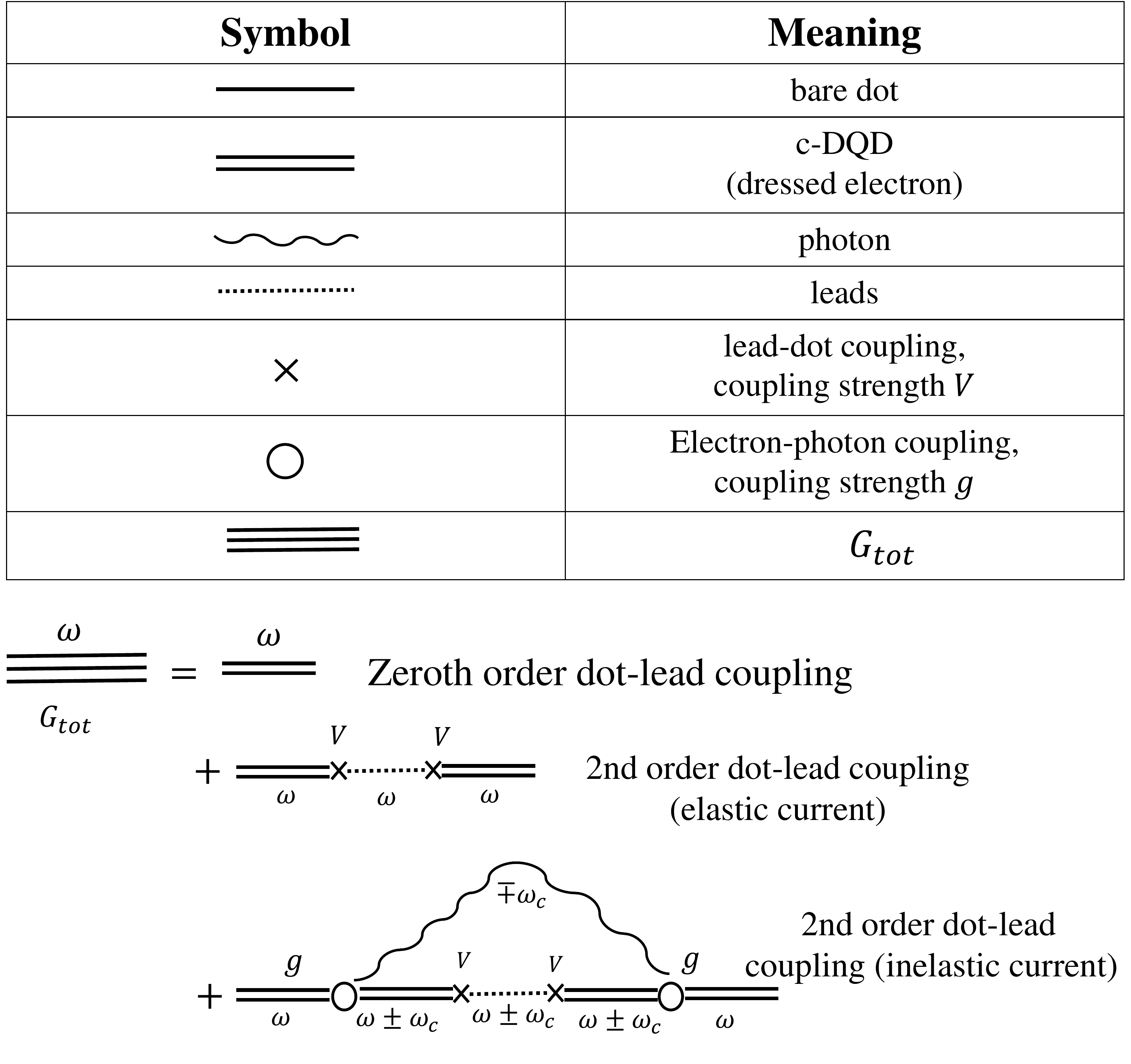}
\caption{(color online).Symbol and Feynman diagram for the Green's function $G_{tot}$ which is the main ingreadient in the transport calculations.}
\label{second}
\end{center}
\end{figure}

\begin{figure}
\begin{center}
\includegraphics[width=3.00 in,height=2.85 in]{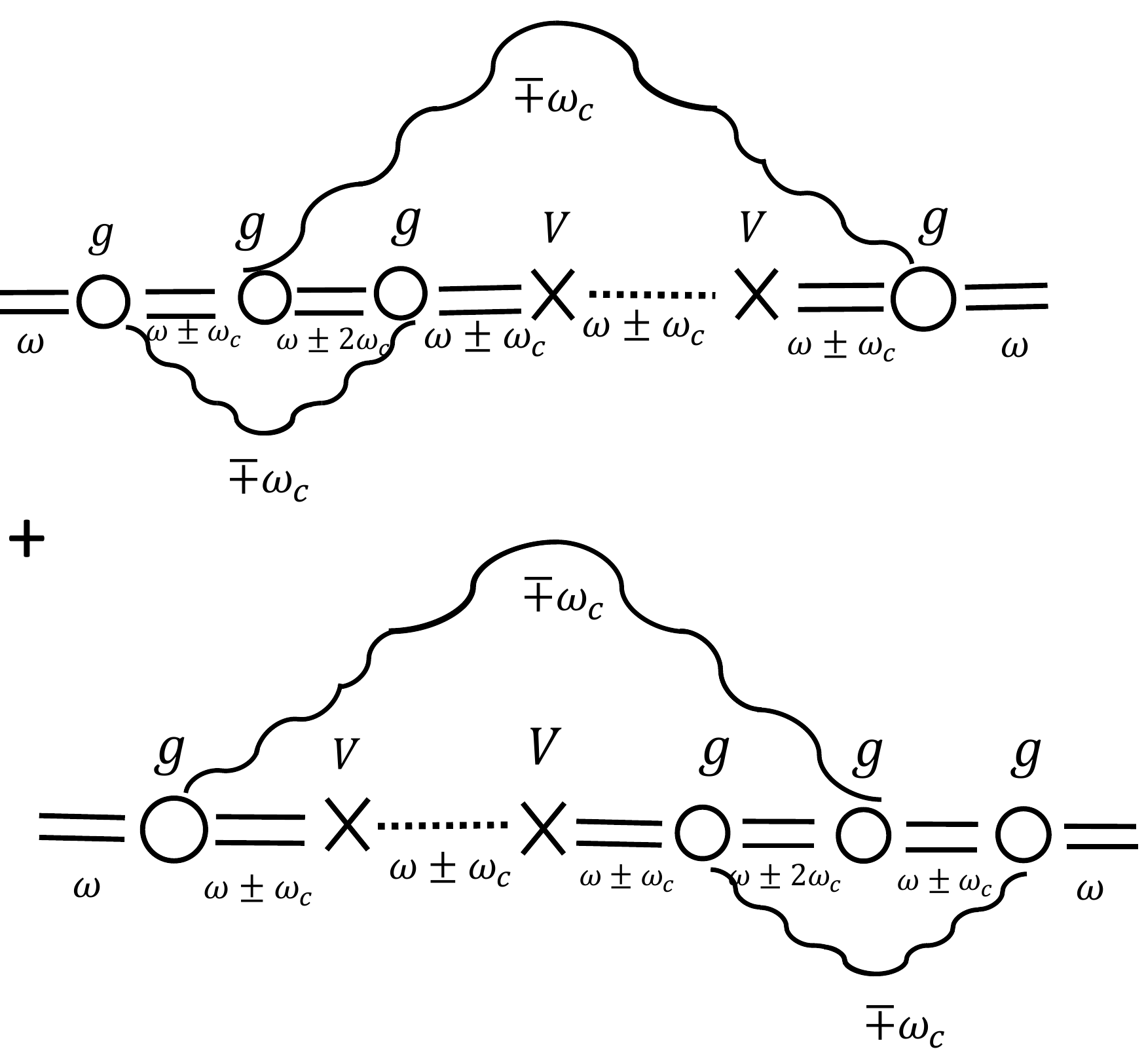}
\caption{(color online).Symbol and Feynman diagram for the Green's function $G_{tot}$ for the 4th order electron-photon interaction.}
\label{forth}
\end{center}
\end{figure}

\section{Enhanced perturbation theory for transport with dot-lead coupling: inelastic and elastic transport currents}
\label{currapp}
With the Green's functions, we can now study the quantum transport by calculating the charge current
\begin{equation}
\begin{aligned}
I_e^{L}=e\frac{d}{dt}\langle \sum_{k}\hat d_i^\dag\hat d_i\rangle=e\int\frac{d\omega}{2\pi}I_{L}(\omega)
\end{aligned}
\end{equation}
and heat current
\begin{equation}
\begin{aligned}
I_Q^{L}=\frac{d}{dt}\langle \sum_{k}(\epsilon_k^{L}-\mu_{L})\hat d_i^\dag\hat d_i\rangle=\int\frac{d\omega}{2\pi}(\omega-\mu_L)I_{L}(\omega)
\end{aligned}
\end{equation}
leaving electrode $L$, The Green's function calculation yields
\begin{align}
& I_{L}(\omega)= -i{\rm Tr}(\hat \Gamma^L_{rot}(\omega)\hat G_{tot}^<(\omega) \nonumber \\
& \hspace{1cm} - f_L(\omega)[\hat G_{tot}^a(\omega)-\hat G_{tot}^r(\omega)]),
\end{align}
which in terms of the total Green's functions are $\hat G_{tot}^<(\omega)$, $\hat G_{tot}^r(\omega)$, and $\hat G_{tot}^a(\omega)$ are the lesser, advanced and retarded Green's function, respectively.
By using the Dyson equation and the Keldysh formula, we have the total retarded (advanced) Green's function,
\begin{equation}
\begin{aligned}
\hat G_{tot}^<(\omega) = \hat G_{tot}^a(\omega)[\hat\Sigma_P^{<}(\omega)+\hat\Sigma_l^{<}(\omega)]\hat G_{tot}^r(\omega),
\end{aligned}
\end{equation}
where
\begin{equation}
\begin{aligned}
\hat G_{tot}^r(\omega)=[(\hat G_{1}^r(\omega))^{-1}-\hat\Sigma_P^r(\omega)]^{-1},
\end{aligned}
\end{equation}
here
\begin{equation}
\begin{aligned}
\hat G_{1}^r(\omega)=[(\hat G_{0}^r(\omega))^{-1}-\hat\Sigma_l^r(\omega)]^{-1},
\end{aligned}
\end{equation}
and
\begin{equation}
\begin{aligned}
\hat G_{0}^r(\omega) = \left( \begin{array}{cccc} G_{0D}^r(\omega) & 0 \\ 0 &
    G_{0d}^r(\omega) \end{array} \right) ,
\end{aligned}
\end{equation}
As is seen from the above equations, the self-energy on the dot includes two contributions. The first, $\Sigma_l$, is due to the coupling with the leads,
\begin{align}
&\hat\Sigma^{>}_l =-i[\hat\Gamma_{rot}^L(1-f_L)+\hat\Gamma_{rot}^R(1-f_R)], \\
&\hat\Sigma^{<}_l =i(\hat\Gamma_{rot}^Lf_L+\hat\Gamma_{rot}^Rf_R), \\
&\hat\Sigma^{r/a}_l =\mp i(\hat\Gamma_{rot}^L+\hat\Gamma_{rot}^R)/2.
\end{align}
The second contribution to the self-energy results from the interaction with the photons, and in the non-crossing approximation (i.e., the
correction of the quantum-dot Green's function due to light-matter interaction is not crossing with the correction due to the dot-lead
coupling) the leading order term is given by,
\begin{align}
\hat\Sigma_P^{r}(\omega)
=&ig^{2}\int\frac{d\omega '}{2\pi}\Bigl [\frac{(1+N_P){\hat G}^{>}_{1}(\omega ')-N_P{\hat G}^{<}_{1}(\omega ')}{\omega -\omega_{c}-\omega '+ i 0^{+}}\nonumber\\
&+\frac{N_P{\hat G}^{>}_{1}(\omega ')-(1+N_P){\hat G}^{<}_{1}(\omega ')}{\omega +\omega_{c}-\omega '+ i 0^{+}}\Bigr ],\
\end{align}
and
\begin{align}
\hat\Sigma^{<}_P(\omega) = g^{2}\Bigl [ N_P{\hat G}^{<}_{1}(\omega -\omega_{c})+(1+N_P) {\hat G}^{<}_{1}(\omega +\omega_{c})\Bigr ],
\end{align}
The leading dependence of the above self-energies on the light-matter interaction is proportional to $g^2$. Nevertheless,
higher-order contributions are also included because of the use of the polaron Green's function.
Inserting the expressions for the Green's function $\hat G_{tot}$ into the above equation, one finds that $I_L$ can be written as a sum of two terms, one arising from the elastic transitions of the transport electrons and the other coming from the inelastic ones,
\begin{align}
I_L(\omega)=I_L^{el}(\omega)+I_L^{inel}(\omega),
\end{align}
The elastic-process contribution is
\begin{align}
& I_L^{el}(\omega)={\rm Tr}(\hat \Gamma^L_{rot}(\omega)\hat G_{tot}^r(\omega)[\hat \Sigma_l^<(\omega)\nonumber \\
& \hspace{1cm} + 2f_L(\omega)\hat \Sigma_l^r(\omega)]\hat G_{tot}^a(\omega)),
\end{align}
while the inelastic one is
\begin{align}
& I_L^{inel}(\omega)={\rm Tr}(\hat \Gamma^L_{rot}(\omega)\hat G_{1}^r(\omega)[\hat \Sigma_P^<(\omega)\nonumber \\
& \hspace{1cm} + 2f_L(\omega)\hat \Sigma_P^r(\omega)]\hat G_{1}^a(\omega)).
\end{align}
So we can get the elastic and inelastic currents,
\begin{subequations}
\begin{align}
& I_e^L|_{el} = e\int \frac{d\omega}{2\pi} {\rm Tr}(\hat \Gamma^L_{rot}(\omega)\hat G_{tot}^r(\omega)[\hat \Sigma_l^<(\omega) \nonumber \\
& \hspace{1cm} + 2f_L(\omega)\hat \Sigma_l^r(\omega)]\hat G_{tot}^a(\omega)) ,\\
& I_e^L|_{inel} =  e\int \frac{d\omega}{2\pi} {\rm Tr}(\hat \Gamma^L_{rot}(\omega)\hat G_{1}^r(\omega)[\hat \Sigma_P^<(\omega) \nonumber \\
& \hspace{1cm} + 2f_L(\omega)\hat \Sigma_P^r(\omega)]\hat G_{1}^a(\omega)).
\end{align}
\end{subequations}
Meanwhile, we obtain the heat current as
\begin{subequations}
\begin{align}
& I_Q^L|_{el} = \int \frac{d\omega}{2\pi} (\omega-\mu_L){\rm Tr}(\hat \Gamma^L_{rot}(\omega)\hat G_{tot}^r(\omega)[\hat \Sigma_l^<(\omega)\nonumber \\
& \hspace{1cm} + 2f_L(\omega)\hat \Sigma_l^r(\omega)]\hat G_{tot}^a(\omega)), \\
& I_Q^L|_{inel} = \int \frac{d\omega}{2\pi} (\omega-\mu_L){\rm Tr}(\hat \Gamma^L_{rot}(\omega)\hat G_{1}^r(\omega)[\hat \Sigma_P^<(\omega)\nonumber \\
&\hspace{1.2cm} + 2f_L(\omega)\hat \Sigma_P^r(\omega)]\hat G_{1}^a(\omega)).
\end{align}
\end{subequations}
We remark that the elastic transport is treated to the second-order in dot-lead coupling and to all orders in light-matter interaction. In contrast, using the enhanced perturbation theory, the inelastic transport is treated at least to the second-order in dot-lead coupling, since the non-perturbative Green's functions of electrons without dot-lead coupling are used (see Fig.~\ref{second}). However, there are still high-order terms missing in our theory (the crossing terms are shown in Fig.~\ref{forth} indicating $g^4$ and higher-order dependences; The non-crossing terms, which are not shown, have at least $g^4$ dependences which correspond to two-photon absorption and emission and other higher-order processes).

\bibliography{Reference}

\end{document}